\documentclass[journal]{IEEEtran}
\usepackage{float}
\usepackage{booktabs}
\usepackage{graphicx}
\usepackage{subfig}
\usepackage{amsmath}
\usepackage{enumitem,amssymb}
\usepackage{bbding} 
\usepackage{cite}
\usepackage{indentfirst}
\usepackage{enumerate}
\usepackage{hyperref}
\usepackage{multirow}
\usepackage{makecell}  
\usepackage{nicematrix}
\usepackage[ruled]{algorithm2e}

\makeatletter
\newcommand{\removelatexerror}{\let\@latex@error\@gobble}
\makeatother

\makeatletter
\renewcommand{\maketag@@@}[1]{\hbox{\m@th\normalsize\normalfont#1}}%
\makeatother

\hypersetup{
	colorlinks=true,
	linkcolor=red,
	urlcolor=blue
}

\usepackage{tikz,xcolor}
\definecolor{lime}{HTML}{A6CE39}
\DeclareRobustCommand{\orcidicon}{%
	\begin{tikzpicture}
	\draw[lime, fill=lime] (0,0) 
	circle [radius=0.134] 
	node[white] {{\fontfamily{qag}\selectfont \tiny ID}};    \draw[white, fill=white] (-0.0625,0.095) 
	circle [radius=0.007];    \end{tikzpicture}
	\hspace{-2mm}}
\foreach \x in {A, ..., Z}{%
	\expandafter\xdef\csname orcid\x\endcsname{\noexpand\href{https://orcid.org/\csname orcidauthor\x\endcsname}{\noexpand\orcidicon}}
}

\begin{document}
	
\captionsetup[figure]{name={Fig.},labelsep=period}
\captionsetup[table]{name={TABLE},labelsep=period}

\title{EDiffSR: An Efficient Diffusion Probabilistic Model for Remote Sensing Image Super-Resolution}

\author{Yi~Xiao\orcidA{},
       Qiangqiang~Yuan\orcidB{},~\IEEEmembership{Member,~IEEE,}
       Kui~Jiang\orcidC{},~\IEEEmembership{Member,~IEEE,} 
       \\Jiang~He\orcidD{},~\IEEEmembership{Graduate Student Member,~IEEE,}
	   Xianyu~Jin\orcidF{}, 
       and~Liangpei~Zhang\orcidE{}, ~\IEEEmembership{Fellow,~IEEE.} 
        
\thanks{This work was supported  in part by the National Natural
	Science Foundation of China under Grant 42230108 and 61971319. (\emph{Corresponding author: Qiangqiang Yuan}.)}
\thanks{Yi Xiao, Qiangqiang Yuan, Jiang He, and Xianyu Jin are with the School of Geodesy and Geomatics, Wuhan University, Wuhan 430000, China (e-mail: xiao\_yi@whu.edu.cn; yqiang86@gmail.com; jiang\_he@whu.edu.cn; jin\_xy@whu.edu.cn).}
\thanks{Kui Jiang is with the School of Computer Science and Technology, Harbin Institute of Technology, Harbin 150000, China (e-mail: kuijiang\_1994@163.com).}
\thanks{Liangpei Zhang is with the State Key Laboratory of Information Engineering in Surveying, Mapping, and Remote Sensing, Wuhan University, Wuhan 430000, China (e-mail: zlp62@whu.edu.cn).}}

%
%

\markboth{Submitted to IEEE Transactions on Geoscience and Remote Sensing}%
{Shell \MakeLowercase{\textit{et al.}}: Bare Demo of IEEEtran.cls for IEEE Journals}
%



\maketitle

\begin{abstract}
Recently, convolutional networks have achieved remarkable development in remote sensing image Super-Resoltuion (SR) by minimizing the regression objectives, \emph{e.g.,} MSE loss. However, despite achieving impressive performance, these methods often suffer from poor visual quality with over-smooth issues. Generative adversarial networks have the potential to infer intricate details, but they are easy to collapse, resulting in undesirable artifacts. To mitigate these issues, in this paper, we first introduce Diffusion Probabilistic Model (DPM) for efficient remote sensing image SR, dubbed EDiffSR. EDiffSR is easy to train and maintains the merits of DPM in generating perceptual-pleasant images. Specifically, different from previous works using heavy UNet for noise prediction, we develop an Efficient Activation Network (EANet) to achieve favorable noise prediction performance by simplified channel attention and simple gate operation, which dramatically reduces the computational budget. Moreover, to introduce more valuable prior knowledge into the proposed EDiffSR, a practical Conditional Prior Enhancement Module (CPEM) is developed to help extract an enriched condition. Unlike most DPM-based SR models that directly generate conditions by amplifying LR images, the proposed CPEM helps to retain more informative cues for accurate SR. Extensive experiments on four remote sensing datasets demonstrate that EDiffSR can restore visual-pleasant images on simulated and real-world remote sensing images, both quantitatively and qualitatively. The code of EDiffSR will be available at \url{https://github.com/XY-boy/EDiffSR}
\end{abstract}


%
\IEEEpeerreviewmaketitle

\begin{figure}[t]
\centering
\includegraphics[width=3.5in]{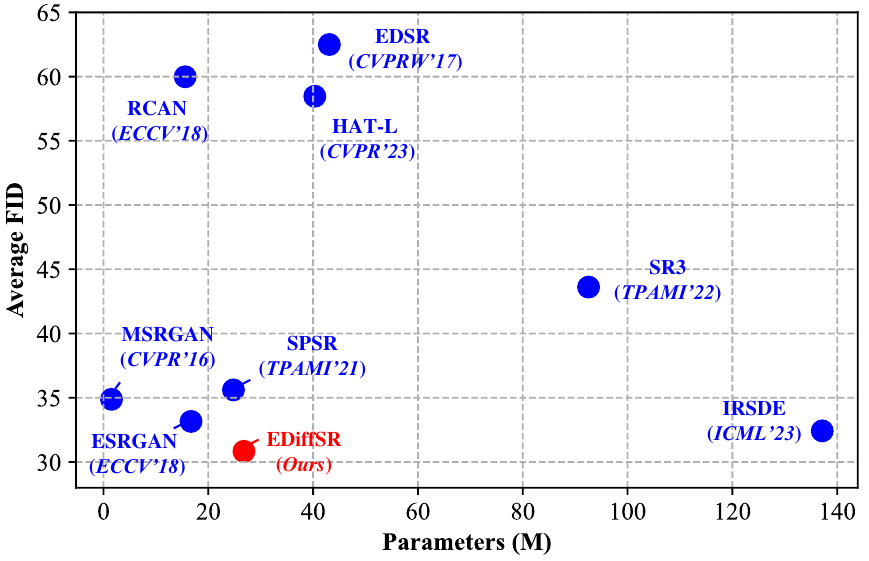}%
\captionsetup{font={scriptsize}}   
\caption{The relationship between FID (Fréchet Inception Distance)~\cite{fid} performance and parameter of state-of-the-art (SOTA) SR methods (lower FID values indicate better generative quality). EDSR~\cite{edsr}, RCAN~\cite{rcan}, and HAT-L~\cite{hat} are regression-based models, typically generating low-quality distribution. The GAN-based approaches (MSRGAN~\cite{srgan}, ESRGAN~\cite{esrgan} and SPSR~\cite{spsr}) and DPM-based methods (SR3~\cite{sr3} and IRSDE~\cite{irsde}) can produce high-quality images. Our EDiffSR achieves the best performance and is far more lightweight than SOTA DPM-based SR models.}
\label{introduction}
\end{figure}

\section{Introduction}

\IEEEPARstart{S}{uper}-Resolution (SR) is a long-standing issue and remains an active research topic in the area of remote sensing \cite{he2023spectral}. SR aims to reconstruct a high-resolution (HR) image with rich texture details from a low-resolution (LR) image \cite{bsvsr}\cite{liu2023efficient}\cite{drw1}. Currently, SR has been widely explored in remote sensing applications, including land-cover mapping \cite{hd1}\cite{hd2}, hyperspectral image fusion \cite{he2023self}\cite{dian2020regularizing}, product reconstruction \cite{wang2022global, xiao2022generating}, and vehicle tracking \cite{xiao2022space}. Due to the inherently ill-posed nature \cite{10144690, miao2021hyperspectral}, SR is challenging because the HR counterpart may be infinite in the solution space given an LR input. In particular, for large-scale earth observation scenarios, SR becomes more complicated owing to various degradations, such as atmospheric scattering and platform tremors. Therefore, developing an effective SR method to reconstruct high-quality images is indispensable and of practical importance.

Convolutional methods have shown significant success in modeling the non-linear relationship between LR and HR images in recent years. 
Among them, various efforts have been made to tame the inherent ill-posedness, such as dense \cite{jiang2020hierarchical,rdn} and residual networks \cite{vdsr}, attention-based models \cite{rcan}\cite{nlsa} and transformer architectures \cite{hat}\cite{he2022dster}. However, existing methods often employ regression function, \emph{e.g.,} Mean Squared Error (MSE) and Mean Absolute Error (MAE), to minimize the pixel-level difference between super-resolved results and ground truth images. Despite obtaining favorable PSNR performance, these optimal objects can lead the model to average the pixel distance, resulting in over-smooth results. To restore visually convincing details, deep Generative Adversarial Networks (GANs)\cite{ledig2017photo} have been explored. Such methods exploit the adversarial optimization between the generator and discriminator to encourage the generator to recover realistic images. Generally, GANs require carefully designed loss functions as auxiliary, \emph{e.g.,} perceptual loss \cite{esrgan} and gradient loss \cite{spsr}, to optimize the distance in the feature domain. Although GANs can generate rich details, they often suffer from training instability and are easy to collapse, leading to undesirable artifacts.

Recently, Diffusion Probabilistic Models (DPM) \cite{ddpm} have received increasing attention in the realm of image-to-image translation, and also achieved promising performance in super-resolution tasks \cite{srdiff}\cite{sr3}\cite{whl}\cite{miao2023dds2m}. The key to DPM is the reverse diffusion process, which iteratively predicts various noises from a noisy image. In this manner, DPM can generate high-quality data distributions from random noise. Thanks to its principled and well-defined probabilistic diffusion process, DPM can mitigate the training instability that commonly occurs in GANs and generate more complex distributions. More recently, Saharia \emph{et al.} \cite{sr3} pioneered the DDPM-based SR method, and utilized the heavy UNet as the denoiser to generate images by iterative refinement. To better simulate the degradation process, Luo \emph{et al.} \cite{irsde} proposed stochastic differential equations to model the diffusion process. Nevertheless, most DPM-based SR methods remain confined within the paradigm of image synthesis tasks, lacking insightful design for SR tasks. Specifically, \textbf{1) The prior knowledge in LR image, which is critical for SR tasks, is rarely explored.} Following the paradigm of image synthesis, the LR image is often directly upsampled by bicubic interpolation to serve as the condition. This pre-processing scheme lacks elaboration and can only convert partial prior knowledge into a diffusion model. As a result, it may lead to suboptimal performance. \textbf{2) The vanilla UNet consumes massive computational cost and is less effective in SR tasks.} In contrast to image synthesis which needs to predict an image from scratch, more pixels of SR are given. Thus, employing a large network for noise prediction is inefficient.

To this end, this paper explores the application of DPM and devises an Efficient Diffusion model for remote sensing image Super-Resolution (EDiffSR). Unlike previous work that applied bicubic-upsampled LR image as the condition, we developed a novel Conditional Prior Enhancement Module (CPEM) to effectively leverage prior knowledge in low-resolution (LR) images. It promotes the condition with more informative and plentiful input. Moreover, an Efficient Activation Block (EAB) is devised to form our denoising network (EANet), achieving favorable denoising capability while maintaining a far more lightweight design. As illustrated in Fig. \ref{introduction}, our EDiffSR achieves impressive performance while with significantly fewer parameters than previous DPM-based SR approaches (\emph{e.g.,} SR3 \cite{sr3} and IRSDE \cite{irsde}). In addition, we equip our EDiffSR with the stochastic differential equations (SDE) \cite{irsde} to further facilitate the sampling process in the diffusion process. Extensive evaluations on four remote sensing datasets demonstrate the superiority of our EDiffSR in both perceptual quality and quantitative metrics over the state-of-the-art CNN-based, GAN-based, and DPM-based models, while with considerable competitiveness in terms of model efficiency.

To sum up, the main contributions of this study are summarized as follows.
\begin{description}
\item 1) We pioneer an efficient yet effective diffusion probabilistic model (EDiffSR) for remote sensing image super-resolution. By introducing more prior knowledge into the diffusion model with elaborate CPEM, our EDiffSR can achieve accurate SR performance.

\item 2) The proposed EANet can exceed the previous SOTA methods in noise prediction while with lower computational cost. It provides a new perspective for exploring more efficient diffusion-based frameworks.
\end{description}

The remainder of this paper is organized as follows. Section \ref{related} reviews the progress of image super-resolution and diffusion models. Section \ref{pre} presents some preliminary of the diffusion process. Section \ref{meth} introduces the implementation details of our EDiffSR. Section \ref{exp} contains experiments and analysis. Section \ref{conclu} is the conclusion.

\section{Related Work}\label{related}
\subsection{Deep Learning-based Image Super-Resolution}

\subsubsection{CNN-based Models}
Inspired by the success of SRCNN \cite{srcnn}, numerous elaborate CNN architectures have been proposed, such as very deep \cite{vdsr} and wide \cite{edsr} architectures, attention mechanism \cite{rcan}\cite{nlsa}, and transformer \cite{hat}. Recently, Chen \emph{et al.} \cite{hat} proposed an impressive method by combining the advantage of channel attention and self-attention to activate more useful information for SR. However, most CNN-based SR approaches predict the target distribution by minimizing the MSE ($L_2$) or MAE ($L_1$) loss. While achieving high PSNR values, the regression functions often tend to encourage the network to "average" a certain region, leading to an undesirable over-smooth issue. In contrast, our EDiffSR can benefit from the generative capability of DPM to recover more realistic distributions, improving the visual quality of SR results by a large margin.

\begin{figure*}[t]
\centering
\includegraphics[width=7.1in]{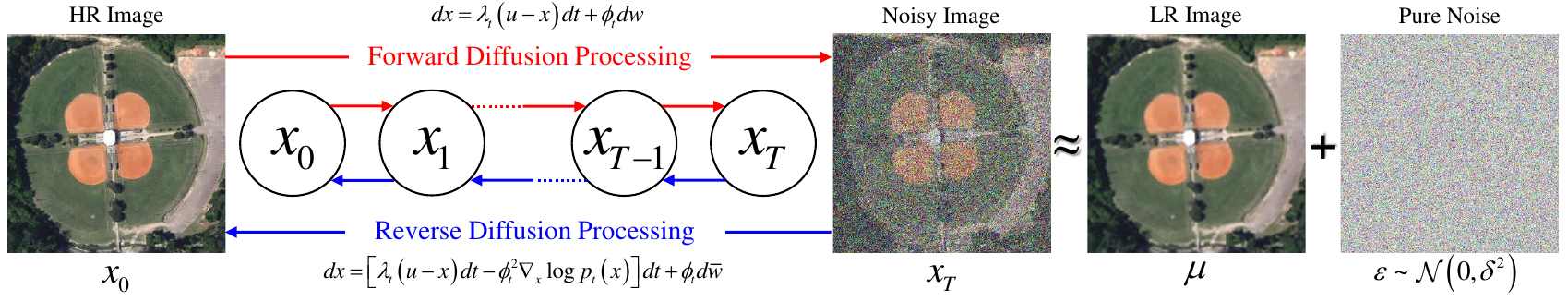}%
\captionsetup{font={scriptsize}}   
\caption{Overview of the forward and reverse diffusion processes defined by mean-reverting Stochastic Differential Equations (SDE). The forward diffusion gradually degrades the high-quality and high-resolution image $x_0$ to its low-quality counterpart via $x_T=\mu+\varepsilon$. The reverse diffusion learns to characterize the noise and reconstruct the corresponding high-resolution image.}
\label{sde}
\end{figure*}

\subsubsection{GAN-based Models}
To promote visual pleasure, GAN-based SR approaches introduce elaborate auxiliary loss to guide the network to generate photo-realistic results. For example, Ledig \emph{et al.} \cite{srgan} pioneered the perceptual loss that measures the feature-wise distance between the restored image and ground-truth image in VGG feature space. To improve the training stability, Wang \emph{et al.} \cite{esrgan} put forward an Enhanced SRGAN (ESRGAN) with modified discriminative constraints while removing batch normalization (BN) to avoid artifacts. Sajjadi \emph{et al.} \cite{enet} developed a texture loss to preserve the high-frequency textual details. Recently, Ma \emph{et al.} \cite{spsr} proposed a structural preserving GAN that maintains structural information by the gradient loss. 
Although GANs can bring impressive improvement in visual quality, they often face harsh optimization problems. Moreover, we often require laborious tricks to strike the balance between these carefully designed loss functions. Benefiting from the well-defined diffusion process, the proposed EDiffSR offers a stable and interpretable training process.

\subsubsection{Diffusion-based Models}
Diffusion models use a fixed Markov chain to optimize the variations boundaries of the likelihood function and have recently received increasing attention due to their excellent performance on generative tasks \cite{rombach2022high}. In SR task, research on diffusion modeling is still in its infancy. Until recently, Saharia \emph{et al.} \cite{sr3} proposed to generate results that exceed those of the GAN with iterative refinement. Li \emph{et al.} \cite{srdiff} firstly introduced the residual prediction in DPM for face image SR. More recently, Xia \emph{et al.} \cite{xia2023diffir} exploited the transformer block to model the long-range discrepancy for effective image restoration. Luo \emph{et al.} \cite{irsde} proposed an averaging-equation idea to simulate the image degradation process while realizing a faster diffusion process. However, current DPM-based SR models often rely on large models for noise prediction. The high complexity of denoisers limits their practical application and leads to inefficient inference in large-scale remote sensing scenarios. In contrast, the proposed EDiffSR achieves favorable noise prediction with a far more lightweight EANet. Besides, existing methods barely consider the prior information in images, which is crucial for SR tasks, thus resulting in suboptimal performance. The proposed EDiffSR seeks a more desirable condition by exploring informative cues from the LR images, which further boosts the SR performance.

\subsection{Remote Sensing Image Super-Resoltuion}
Early SR methods are CNN-based, aiming to achieve high PSNR performance \cite{jiang2018deep}\cite{lei2021transformer}\cite{d2u}. In this parse, more efforts have been paid to improve the network structure, making the convolution network grasp more characteristics of remote sensing images, such as scene-adaptive network \cite{zs}, multi-scale \cite{msdtgp}\cite{hsenet}\cite{chenshi} and multi-stage \cite{lei2021transformer} design, numerous attentions \cite{mhan}\cite{lgtd} and guided super-resolution \cite{yuanyuan1}. Li \emph{et al.} \cite{yuanyuan3} proposed a novel dual-stage network to reconstruct more missing details in remote sensing imagery in a coarse-to-fine manner. Recently, Li \emph{et al.} \cite{yuanyuan2} put forward to transfer more beneficial supplementary from RGB images to remote sensing scenes. However, as PSNR tends to penalize the reconstruction of high-frequency details, these methods can not reflect human preference well in RSI.

To recover rich detailed information in RSI, various GAN-based methods have been proposed. Lei et al. \cite{lei2019coupled} a coupled-discriminated GAN to make better discrimination. Jiang \emph{et al.} \cite{jiang2019edge} proposed an edge-enhanced GAN by optimizing the high-frequency and low-frequency components simultaneously. Haut \emph{et al.} \cite{usrgan} proposed to train a GAN in an unsupervised manner without the HR RSI. Recently, Tu \emph{et al.} \cite{tu2022swcgan} incorporated the long-range modeling capability of the Swin transformer into GAN, achieving favorable perceptual quality of SR results. However, these approaches often involve complex optimization functions and network structures, leading to training instability. In contrast, this paper proposed an efficient solution to recover perceptual-pleasant RSI, mitigating the training instability of GAN.

In the area of remote sensing, some researchers have applied diffusion modeling to SR tasks \cite{liu2022diffusion}\cite{han2023enhancing}\cite{whl}. However, they borrow too much from the paradigm in image synthesis, which uses a large UNet for noise estimation, resulting in inefficient inference in SR tasks. In addition, there is a lack of consideration of incorporating the valuable prior knowledge in diffusion to generate high-frequency details in remote sensing images. In this paper, we demonstrate that a low-complexity network can provide a more practical and efficient scheme to deliver competitive denoising performance in SR tasks when compared to state-of-the-art methods employing larger models like UNet. In addition, unlike simple bicubic upsampling, we choose to explore more prior information to generate informative conditions, thus further enhancing the diffusion model to generate realistic distributions.

\begin{figure*}[!t]
\centering
\includegraphics[width=7.1in]{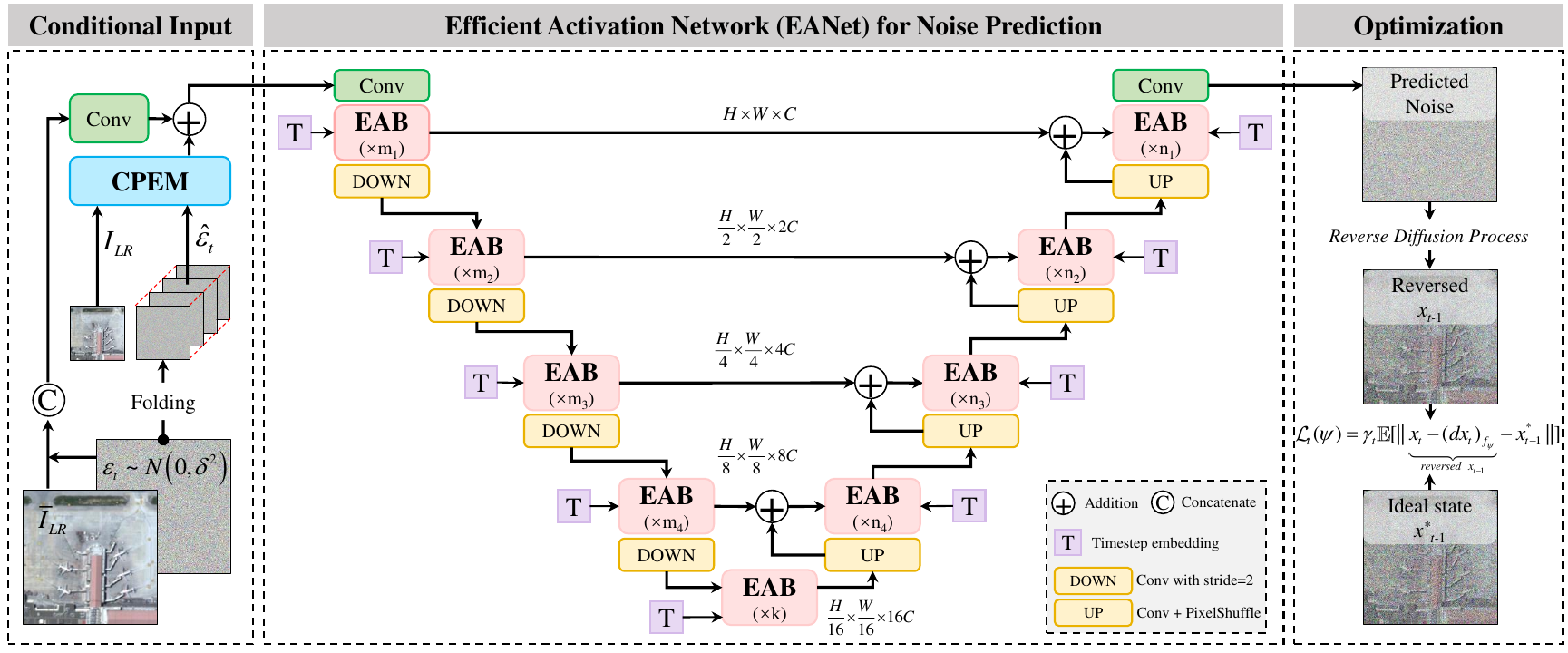}%
\captionsetup{font={scriptsize}}   
\caption{Overall framework of our EDiffSR. It consists of three parts: the condition part, Efficient Activation Network (EANet), and the optimization part. In the condition part, CPEM is designed to explore more priors from the original LR image. EANet takes the condition as input and characterizes the noise distribution. Compared to the primary Unet, it is more efficient and effective owing to the efficient activation block (EAB). The optimization process adopts the maximum likelihood learning for a more stable diffusion process.}
\label{network}
\end{figure*}

\begin{figure}[!t]
\centering
\includegraphics[width=3.5in]{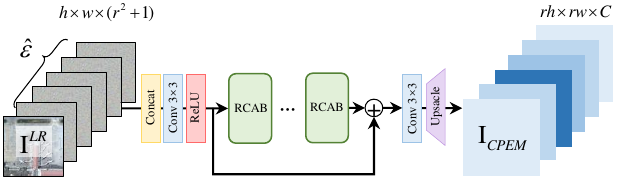}%
\captionsetup{font={scriptsize}}   
\caption{The flowchart of Conditional Prior Enhancement Module (CPEM). We adopt the Residual Channel Attention Block (RCAN) to extract rich prior information.} 
\label{cpem}
\end{figure}

\section{Preliminary}\label{pre}
\subsection{Forward Diffusion Process}
The forward diffusion process aims to gradually transform the initial data distribution $x_0$ to a noisy image $x_T$ after time step $T$. As shown in Fig. \ref{sde}, we define the ground truth image $I_{HR}$ as $x_0$. As such, $x_T$ can approximately to the combination  of the bicubic-upsampled LR image $\mu$ and a pure Gaussian noise $\varepsilon \sim {\rm{{\cal N}}}(0,{\delta ^2})$. Here $\delta ^2$ represents the stationary variance.
This paper adopts the mean-reverting Stochastic Differential Equations (SDE) \cite{irsde} to define the diffusion process as it allows for an efficient sampling process. Specifically, as illustrated in Fig. \ref{sde}, the forward diffusion process is depicted as
\begin{equation}
dx = {\lambda _t}\left( {u - x} \right)dt + {\phi _t}dw, \label{eq1}
\end{equation}
where $w$ refers to a standard Wiener process. $\lambda _t$ and $\rho _t$ are two time-dependent parameters that control the speed of mean reversion and stochastic volatility, respectively. To make equation \ref{eq1} have a closed-form solution, we set $\phi _t^2/\lambda _t = 2{\delta ^2}$. As shown in Fig. \ref{sde}, given an HR image $x_0$ and $t \in [0,T]$, for an intermediate moment $t$, the corresponding state $x_t$ can be strictly expressed by the closed-form solution of Eq. (\ref{eq1}):
\begin{equation}
{x_t} = \mu  + ({x_0} - \mu ){e^{ - {{\bar \lambda }_t}}} + \int_0^t {{\phi _z}{e^{ - {{\bar \lambda }_t}}}} dw(z), \label{eq2}
\end{equation}
where ${\bar \lambda _t}$ is equal to $\int_0^t {{\lambda _z}dz}$. The proof of Eq. \ref{eq2} can be found at \cite{irsde}. In this case, $x_t$ follows a Gaussian probability distribution $p_t(x)$, expressed as
\begin{equation}
{x_t}\sim{p_t}(x) = {\rm{{\cal N}}}({x_t}|{m_t}(x),{n_t})
\end{equation}
where ${m_t(x)} = \mu  + ({x_0} - \mu ){e^{ - {{\bar \lambda }_t}}}$ and $n_t={\delta ^2}(1 - {e^{ - 2{{\bar \lambda }_t}}})$ are the mean and variance of this Gaussian distribution, respectively. It is easy to observe that as the diffusion time $t \to \infty $, $m_t$ and $n_t$ would converge to $\mu$ and ${\delta ^2}$, \emph{i.e.,} the terminal state $x_T \approx \mu + \varepsilon$, which aligns with the aim of the forward diffusion process. 

\subsection{Reverse Diffusion Process}
Reverse diffusion aims to recover the HR image from the terminal state $x_T$. We can define the reverse diffusion process by simulating the reverse-time SDE \cite{anderson1982reverse} as 
\begin{equation}
dx = \left[ {{\lambda _t}\left( {u - x} \right)dt - \phi _t^2{\nabla _x}\log {p_t}\left( x \right)} \right]dt + {\phi _t}d\bar w.\label{rsde}
\end{equation}
where $\bar w$ denotes a reverse-time Wiener process. ${\nabla _x}\log {p_t}\left( x \right)$ is the ground-truth score during inference stage. Note that in the training stage, the ground-truth image $x_0$ is available, thus we can leverage more pleasurable conditional scores during model training. In particular, it can be defined by
\begin{equation}
{\nabla _x}\log {p_t}(x|{x_0}) =  - \frac{{{x_t} - {m_t}(x)}}{{{n_t}}}.
 \end{equation}
Furthermore, if we re-parameterize $x_t$ to ${x_t} = {m_t}(x) + \sqrt {{n_t}} {\varepsilon _t}$, where $\varepsilon _t$ is a standard Gaussian noise 
with the distribution ${\rm{{\cal N}}}(0,I)$. The ground-truth scores can be expressed as $- \frac{{{\varepsilon _t}}}{{\sqrt {{n_t}} }}$. Since $m_t(x)$ and $n_t$ are known, then we just need to estimate the noise using a noise prediction network ${f_\psi }$.

Similar to DDPM \cite{ddpm}, we compute the Euclidean distance between the predicted noise and ground truth noise $\varepsilon _t$ by the following formula:
\begin{equation}
{{\rm{{\cal L}}}}(\psi ) = \sum\nolimits_{t = 0}^T {{\gamma _t}\mathbb{E}[||\underbrace {{f_\psi }\left( {{x_t},u,v,t} \right)}_{predicted~noise~{{\bar \varepsilon }_t}} - {\varepsilon _t}||]}, \label{eq6}
\end{equation}
where $\gamma _t$ denotes the positive weight, and $v$ refers to the original LR image.

\section{Proposed Method}\label{meth}

\subsection{Overview}
Fig. \ref{network} details the flowchart of our proposed EDiffSR. In the input part, we perform conditional prior enhancement to generate a more pleasurable condition for noise prediction. Specifically, the prior enhancement module $f_{CPEM}$ takes the random noise $\varepsilon_t$, LR image $v$, and the corresponding bicubic-upsampled LR image ${\bar I_{LR}}$ as input, and then produce the enriched condition ${\mathop{\rm I}\nolimits} _t$ by the following formula:
\begin{equation}
{{\mathop{\rm I}\nolimits} _t} = {f_{CPEM}}(v,{{\hat \varepsilon }_t}) + {\mathop{\rm f_3}\nolimits} ([\mu ,{\varepsilon _t}]),\label{eq7}
\end{equation}
where ${\hat \varepsilon _t} = {\mathop{\rm Fold}\nolimits} ({\varepsilon _t})$ represents that we adopt the pixel-folding operator to downsample the scale of $\varepsilon _t$ without loss spatial information. ${\mathop{\rm f_3(\cdot)}\nolimits}$ is a $3\times3$ convolution, and $[\cdot]$ represents channel-wise concatenation.
Subsequently, a conditional time-dependent network ${{f_\psi }}$  takes the pleasurable condition and time $t$ as input, aiming to output a pure noise:
\begin{equation}
{\bar \varepsilon _t} = {f_\psi }\left( {{{\mathop{\rm I}\nolimits} _t},t} \right).
\end{equation} 
Here, we adopt the efficient activation network (EANet) for noise prediction. Finally, we can optimize $f_\psi$ until it converges.

\begin{figure}[t]
\centering
\includegraphics[width=3.3in]{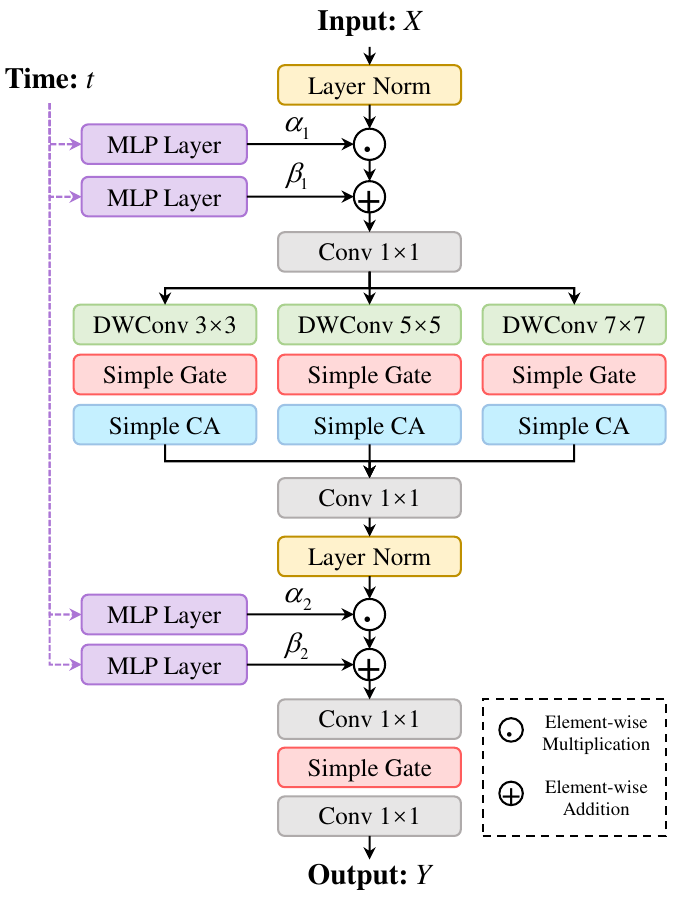}%
\captionsetup{font={scriptsize}}   
\caption{Illustration of the Efficient Activation Block (EAB). } 
\label{eab}
\end{figure}

\subsection{Conditional Prior Enhancement Module}
Most previous methods have typically prepared the condition input by simply upsampling the LR image using bicubic interpolation. However, for SR tasks, this scheme may lose critical structure information, resulting in suboptimal conditional inputs. In contrast, our EDiffSR proposes to generate a more informative condition by exploring additional prior knowledge from the LR image, thus enriching the condition information for better SR performance.

As shown in Fig. \ref{cpem}, the CPEM mainly consists of a convolution layer and ReLU activation, followed by stacked Residual Channel Attention Blocks (RCABs) \cite{rcan} and an upscale layer. To unify the scale of noise and the LR image, we first convert $\varepsilon_t$ to the noisy cube using pixel folding. Then we concatenate and pass them through a $3\times3$ convolution layer followed by ReLU activation to perform shallow feature extraction, depicted as
\begin{equation}
{{\mathop{\rm I}\nolimits} _0} = {\mathop{\rm ReLU}\nolimits} ({\mathop{\rm Conv}\nolimits} ({\mathop{\rm Concat}\nolimits} (v,\hat \varepsilon_t ))).
\end{equation}
Subsequently, $n$ cascaded RCABs $f_{RCAB}$ are used to achieve deep feature extraction and stabilize the gradient by global residual connection. 
\begin{equation}
{{\mathop{\rm I}\nolimits} _{deep}} = {f_{RCAN} ^n}({{\mathop{\rm I}\nolimits} _0}) + {{\mathop{\rm I}\nolimits} _0}.
\end{equation}
Following that, a $3\times3$ convolution and a ${\mathop{\rm PixelShuffle}\nolimits}$ layer \cite{espcn} are used to get the condition yield from CPEM.
\begin{equation}
{{\mathop{\rm I}\nolimits} _{CPEM}} = {\mathop{\rm PixelShuffle}\nolimits} ({\mathop{\rm Conv}\nolimits} ({{\mathop{\rm I}\nolimits} _{deep}})).
\end{equation}
According to Eq. (\ref{eq7}), we generate the improved condition.

\subsection{Efficient Activation Network for Noise Prediction}
As illustrated in Fig. \ref{network}, the key component of our EANet is the Efficient Activation Block (EAB). Fig. \ref{eab} displays the architecture of the EAB, showcasing its lightweight design. The EAB primarily consists of Depth-Wise Convolution (DWConv), simple channel attention, and simple gate operations. This lightweight design results in significantly lower computational complexity when compared to large UNet architectures that incorporate channel attention or self-attention mechanisms. As discussed before, in the context of SR tasks, the majority of pixels are known. Therefore, a large model running massive calculations is inefficient in SR and may lead to a suboptimal performance due to redundant inference. Our EANet offers a more practical scheme to achieve favorable denoising performance with a lightweight model.

\begin{figure}[!t]
\centering
\includegraphics[width=3.5in]{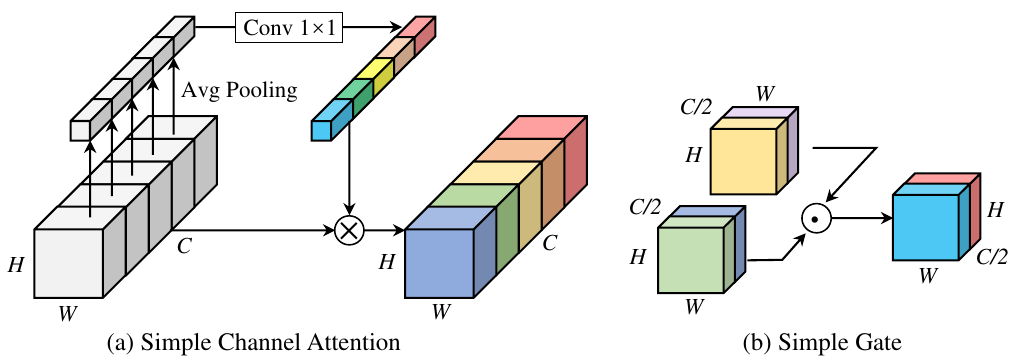}%
\captionsetup{font={scriptsize}}   
\caption{The flowchart of (a) Simple Channel Attention and (b) Simple Gate operation.} 
\label{sca}
\end{figure}

As shown in Fig. \ref{eab}, given an input ${\mathop{\rm X}\nolimits}$ and time step $t$, EAB predict an output ${\mathop{\rm Y}\nolimits}$. In particular, $t$ will be projected to two flatten features by MLP layer for feature modulation:
\begin{equation}
{\mathop{\rm F}\nolimits}  = {f_{1 \times 1}}({\alpha _1} \odot {\mathop{\rm Norm}\nolimits} ({\mathop{\rm X}\nolimits} ) + {\beta _1}).
\end{equation}
Subsequently, we use multi-scale DWConv to explore the multi-scale knowledge in RSI. Within each scale, we incorporate additional nonlinear representations through the use of simple channel attention and simple gate activation, as illustrated in Fig. \ref{sca}. To simplify the channel attention \cite{rcan}, we eliminate the convolution layer and sigmoid activation. The simple gate activation is essentially an element-wise product operation applied to the feature maps. The multi-scale simple activation process can be expressed as
\begin{equation}
\left\{ \begin{array}{l}
{\mathop{\rm F}\nolimits}^3 = {\mathop{\rm SCA}\nolimits} ({\mathop{\rm SimpleGate}\nolimits} ({f_{3 \times 3}}({\mathop{\rm F}\nolimits})))\\
{\mathop{\rm F}\nolimits}^5 = {\mathop{\rm SCA}\nolimits} ({\mathop{\rm SimpleGate}\nolimits} ({f_{5 \times 5}}({\mathop{\rm F}\nolimits})))\\
{\mathop{\rm F}\nolimits}^7 = {\mathop{\rm SCA}\nolimits} ({\mathop{\rm SimpleGate}\nolimits} ({f_{7 \times 7}}({\mathop{\rm F}\nolimits})))
\end{array} \right.
\end{equation}
After that, we set a $1\times1$ convolution to aggregate these multi-scale representations ${\mathop{\rm F}\nolimits}' = {f_{1 \times 1}}({\mathop{\rm Concat}\nolimits} ({\mathop{\rm F}\nolimits}^3,{\mathop{\rm F}\nolimits}^5,{\mathop{\rm F}\nolimits}^7))$. After layer norm, we conduct modulation with the scaling and the shifting operation:
\begin{equation}
{\bar {\mathop{\rm F}\nolimits}} = {\alpha _2} \odot {\mathop{\rm Norm}\nolimits} ({\mathop{\rm F}\nolimits}') + {\beta _2}.
\end{equation}
At the final, the output ${\mathop{\rm Y}\nolimits}$ can be obtained via the following formulation.
\begin{equation}
{\mathop{\rm Y}\nolimits}  = {f_{1 \times 1}}({\mathop{\rm SimpleGate}\nolimits} ({f_{1 \times 1}}({\bar {\mathop{\rm F}\nolimits}})).
\end{equation}

Following previous works \cite{whl, irsde}, we form our EANet to a U-shape encoder-decoder structure. During the encoding phase, we employ a sequence of EABs and a convolution operation with a stride of 2 to progressively downsample the feature maps. In the decoding phase, multiple EABs and pixel-shuffle layers are used to upscale the features. The number of EABs in the encoder and decoder component is denoted as $[{m_1},{m_2},{m_3},{m_4}]$ and $[{n_1},{n_2},{n_3},{n_4}]$, respectively. Additionally, we introduce $k$ EABs in the middle of the EANet.

\begin{figure}[t]
\removelatexerror 
\begin{algorithm*}[H]
\caption{Training of our EDiffSR}
\label{training}
\LinesNumbered 
\KwIn{HR image $x_0=I_{HR}, $LR image $v=I_{LR}$, upsampled LR image $\mu={\bar I_{LR}}$, total step $T$.}

{\bf Initialization:} Random sample $\varepsilon _t\sim {\rm{{\cal N}}}(0, {\delta ^2})$, $t \in [0,T]$, $T=100$.\\

\textbf{repeat}

${{\mathop{\rm I}\nolimits} _t} = {f_{CPEM}}(v,{\hat \varepsilon _t}) + {\mathop{\rm Conv}\nolimits} ([\mu ,{\varepsilon _t}])$;\tcp*[f]{Enhance}\\

${\bar \varepsilon _t} = {f_\psi }\left( {{{\mathop{\rm I}\nolimits} _t},t} \right)$;\tcp*[f]{Predict noise}\\

\tcp*[h]{Substitute score into (6)}\\

$d{x_t} = [{\lambda _t}\left( {u - {x_t}} \right)dt - \phi _t^2\frac{{{{\bar \varepsilon }_t}}}{{\sqrt {{n_t}} }}]dt + {\phi _t}d\bar w$;

${{\rm{{\cal L}}}}(\psi ) = {{\gamma _t}} \mathbb{E}[||\underbrace {{x_t} - {{(d{x_t})}_{{f_\psi }}}}_{reversed\;{x_{t - 1}}} - x_{t - 1}^*||]$;\tcp*[f]{Loss}\\

${\nabla _\psi }{\rm{{\cal L}}}$;\tcp*[f]{Gradient descent}\\

\textbf{until} converged
\end{algorithm*}
\end{figure}

\subsection{Optimization and Inference}
Although Eq. (\ref{eq6}) can provide a straightforward solution to optimize the EANet, the diffusion model often suffers from instability in the training process. The key reason is predicting an instantaneous distribution of noise is not an easy task. Therefore, we modified the training object by using a maximum likelihood learning strategy used in \cite{irsde}. To optimize EANet, specifically, we choose to minimize the Euclidean distance below:
\begin{equation}
{{\rm{{\cal L}}}}(\psi ) = \sum\nolimits_{t = 0}^T {{\gamma _t}} \mathbb{E}[||\underbrace {{x_t} - {{(d{x_t})}_{{f_\psi }}}}_{reversed\;{x_{t - 1}}} - x_{t - 1}^*||],
\end{equation}
where $x_{t - 1}^*$ is the ideal state reversed from $x_t$. The closed-form of $x_{t - 1}^*$ can be determined by the following formula:
\begin{equation}
\begin{split}
x_{t - 1}^* = &\frac{{1 - {e^{ - 2{{\bar \lambda }_{t - 1}}}}}}{{1 - {e^{ - 2{{\bar \lambda }_t}}}}}{e^{ - {{\lambda '}_t}}}({x_t} - \mu ) \\
+ &\frac{{1 - {e^{ - 2{{\lambda '}_t}}}}}{{1 - {e^{ - 2{{\bar \lambda }_t}}}}}{e^{ - {{\bar \lambda }_{t - 1}}}}({x_0} - \mu ) + \mu .
\end{split}
\end{equation}

The proof can be referred to \cite{irsde}. In brief, we transformed the distance between the predicted noise and ground-truth noise into another domain, \emph{i.e.,} the distance between the ideal state and predicted states. This scheme helps to reduce the optimization instability as most pixels in reversed states are known.

In the inference procedure, we utilize the pre-trained $f_\psi$ to predict high-resolution images by sampling from the random state $x_T$ and iteratively solve the SDE with numerical solutions, such as the Euler–Maruyama method \cite{kloeden1992stochastic}. To better understand the training and inference process of our EDiffSR, we summarize these processes in two algorithms, as presented in Algorithm. \ref{training} and Algorithm \ref{inference}.

\section{Experiment and Discussion}\label{exp}
In this section, we conduct extensive experiments on four remote-sensing datasets to evaluate the performance of our EDiffSR, both in simulated and real-world scenarios.
\subsection{Dataset}
We use 4 public remote sensing datasets to comprehensively evaluate the effectiveness of the methods in this paper, including AID \cite{aid}, DOTA \cite{dota}, DIOR \cite{dior}, and NWPU-RESISC45 \cite{nwpu}. The training set in this study consists of 3,000 randomly selected images from the AID dataset with an image size of $640\times640$. Specifically, we randomly select 100 images in each of the 30 categories of AID to build the training set. Additionally, we have selected 10 images from each category that do not overlap with the training set to form the test set, resulting in a total of 300 test images. Furthermore, we have used a subset of images from the DOTA dataset and the DIOR dataset for testing, consisting of 700 and 1,000 images, respectively. These images have a resolution of $512\times512$. As a result, our test set comprises a total of 2,000 images. In our simulated experiments, we used bicubic interpolation for image degradation. NWPU-RESISC45 data were only used for real-world analysis without any simulated degradation. To save the inference cost, we randomly selected 315 images from NWPU and cropped them to $128\times128$.

\begin{figure}[t]
\removelatexerror 
\begin{algorithm*}[H]
\caption{Inference of our EDiffSR}
\label{inference}
\LinesNumbered 
\KwIn{LR image $v=I_{LR}$, upsampled LR image $\mu={\bar I_{LR}}$, total step $T$.}
\KwOut{The super-resolved image $I_{SR}$.}

{\bf Initialization:} Random sample $x_T\sim {\rm{{\cal N}}}(0, {\delta ^2})$, ${{f_\psi }}$ is the pre-trained EANet, ${\mathop{\rm EM}\nolimits}(\cdot)$ is Euler–Maruyama method, $T=100$.\\

\For{ $t = T:1$}{
${{\bar \varepsilon }_t} = {f_\psi }\left( {{x_t},u,v,t} \right)$;\tcp*[f]{Predict noise}\\

\tcp*[h]{Substitute score into (6)}\\

$d{x_t} = [{\lambda _t}\left( {u - {x_t}} \right)dt - \phi _t^2\frac{{{{\bar \varepsilon }_t}}}{{\sqrt {{n_t}} }}]dt + {\phi _t}d\bar w$;

${x_{t - 1}} = {x_t} - {\mathop{\rm EM}\nolimits} (d{x_t})$;\tcp*[f]{Reverse SDE}
}

$I_{SR}=x_0$;

\end{algorithm*}
\end{figure}

\begin{table*}[!t]
  \centering
\captionsetup{font={scriptsize}}   
  \caption{Quantitative FID comparison with state-of-the-art SR models on 30 scene categories of the AID test set. The best FID value in each category is highlighted in \textcolor{red}{\textbf{red}} while the second best is in \textcolor{blue}{\textbf{blue}}.}
\setlength{\tabcolsep}{1.4mm}{
    \begin{tabular}{ccccccccccc}
\toprule[1.5pt]
    Categories & Bicubic & EDSR \cite{edsr}  & RCAN \cite{rcan}  & HAT-L \cite{hat} & MSRGAN \cite{srgan} & ESRGAN \cite{esrgan} & SPSR \cite{spsr} & SR3 \cite{sr3}  & IRSDE \cite{irsde} & EDiffSR \\
\midrule
    Airport & 126.23 & 87.25 & 89.52 & 86.55 & 54.42 & 54.85 & 57.98 & 56.56 & \textcolor{blue}{\textbf{54.27}} & \textcolor{red}{\textbf{52.76}} \\
    Bare Land & 113.49 & 91.50  & 92.51 & 91.15 & 66.83 & \textcolor{red}{\textbf{60.75}} & 72.17 & 76.89 & 80.30  & \textcolor{blue}{\textbf{66.76}} \\
    Baseball Field & 131.28 & 89.17 & 91.09 & 90.06 & \textcolor{blue}{\textbf{51.25}} & \textcolor{red}{\textbf{46.87}} & 55.88 & 70.22 & 57.67 & 52.43 \\
    Beach & 121.31 & 104.78 & 106.03 & 101.27 & 50.90  & 48.96 & 52.78 & 50.81 & \textcolor{blue}{\textbf{43.22}} & \textcolor{red}{\textbf{43.10}} \\
    Bridge & 137.67 & 80.01 & 82.27 & 80.93 & \textcolor{blue}{\textbf{47.43}} & 50.70  & 49.40  & 73.65 & \textcolor{red}{\textbf{45.71}} & 50.98 \\
    Center & 140.22 & 71.42 & 74.34 & 71.55 & 50.27 & 54.30  & 48.24 & 52.00    & \textcolor{blue}{\textbf{44.29}} & \textcolor{red}{\textbf{43.13}} \\
    Church & 122.26 & 85.76 & 87.92 & 89.48 & 51.88 & 51.89 & 55.03 & 62.58 & \textcolor{blue}{\textbf{50.76}} & \textcolor{red}{\textbf{50.70}} \\
    Commercial & 112.45 & 109.99 & 110.88 & 104.21 & 55.42 & 56.18 & 60.77 & 69.68 & \textcolor{red}{\textbf{51.84}} & \textcolor{blue}{\textbf{55.20}} \\
    DenseResidential & 126.36 & 113.85 & 125.26 & 118.28 & 52.16 & 57.75 & 55.99 & 62.96 & \textcolor{red}{\textbf{39.53}} & \textcolor{blue}{\textbf{39.97}} \\
    Desert & 115.10 & 77.50  & 76.95 & 75.84 & 56.28 & \textcolor{blue}{\textbf{53.54}} & 64.02 & 59.35 & 61.50  & \textcolor{red}{\textbf{52.91}} \\
    Farmland & 144.78 & 92.15 & 93.80  & 96.13 & 66.34 & \textcolor{blue}{\textbf{56.12}} & 58.00    & 79.39 & 61.85 & \textcolor{red}{\textbf{51.07}} \\
    Forest & 103.57 & 88.79 & 93.45 & 96.26 & 59.69 & 64.36 & 62.01 & 72.07 & \textcolor{blue}{\textbf{48.68}} & \textcolor{red}{\textbf{46.90}} \\
    Industrial & 106.82 & 77.84 & 80.85 & 74.83 & 37.79 & \textcolor{blue}{\textbf{37.11}} & 45.90  & 46.07 & \textcolor{red}{\textbf{35.90}}  & 41.27 \\
    Meadow & 133.81 & 107.78 & 106.18 & 103.1 & 95.57 & 68.95 & \textcolor{red}{\textbf{64.83}} & 87.56 & 70.93 & \textcolor{blue}{\textbf{66.53}} \\
    MediumResidential & 117.19 & 98.74 & 104.17 & 100.24 & 46.17 & 50.11 & 49.75 & 73.45 & \textcolor{blue}{\textbf{41.80}}  & \textcolor{red}{\textbf{40.05}} \\
    Mountain & 103.15 & 105.5 & 105.64 & 103.68 & 57.93 & \textcolor{blue}{\textbf{54.93}} & 71.01 & 72.67 & 59.02 & \textcolor{red}{\textbf{52.21}} \\
    Park  & 137.79 & 109.64 & 112.28 & 109.54 & \textcolor{red}{\textbf{61.02}} & \textcolor{red}{\textbf{60.33}} & 72.14 & 80.81 & 63.78 & 63.29 \\
    Parking & 134.86 & 60.79 & 67.40  & 63.98 & 41.79 & 42.99 & 45.40  & 56.09 & \textcolor{red}{\textbf{37.02}} & \textcolor{blue}{\textbf{36.74}} \\
    Playground & 113.86 & 58.07 & 61.90  & 60.36 & 40.69 & 39.15 & 41.00  & 53.96 & \textcolor{blue}{\textbf{38.89}} & \textcolor{red}{\textbf{35.94}} \\
    Pond  & 162.50 & 122.29 & 124.27 & 126.31 & 60.65 & \textcolor{red}{\textbf{54.71}} & 62.62 & 104.36 & 56.29 & \textcolor{blue}{\textbf{55.97}} \\
    Port  & 134.94 & 77.12 & 80.02 & 80.55 & 46.76 & \textcolor{blue}{\textbf{46.72}} & 52.02 & 58.93 & \textcolor{red}{\textbf{46.52}} & 48.22 \\
    Railway Station & 113.35 & 93.26 & 93.77 & 87.06 & 50.38 & 52.08 & 58.44 & 56.59 & \textcolor{red}{\textbf{49.82}} & \textcolor{blue}{\textbf{51.89}} \\
    Resort & 131.05 & 99.11 & 104.87 & 105.46 & 59.79 & 61.77 & 67.71 & 69.35 & \textcolor{blue}{\textbf{59.00}}    & \textcolor{red}{\textbf{57.26}} \\
    River & 151.14 & 106.24 & 109.20 & 108.06 & \textcolor{red}{\textbf{54.50}}  & 59.23 & 65.18 & 83.28 & 60.27 & \textcolor{blue}{\textbf{57.34}} \\
    School & 110.22 & 85.48 & 89.16 & 82.25 & 49.53 & 50.33 & 53.65 & 60.20  & \textcolor{blue}{\textbf{47.60}}  & \textcolor{red}{\textbf{47.00}} \\
    SparseResidential & 149.02 & 134.24 & 140.41 & 132.73 & 73.57 & 75.55 & 77.83 & 85.06 & \textcolor{red}{\textbf{69.59}} & \textcolor{blue}{\textbf{71.52}} \\
    Square & 108.42 & 70.79 & 75.52 & 71.89 & \textcolor{blue}{\textbf{42.48}} & 44.89 & 46.29 & 52.92 & 45.04 & \textcolor{red}{\textbf{42.43}} \\
    Stadium & 121.79 & 56.48 & 59.39 & 58.87 & 37.70  & 35.28 & 37.42 & 38.27 & \textcolor{red}{\textbf{32.59}} & \textcolor{blue}{\textbf{34.18}} \\
    Storage Tanks & 161.44 & 89.80  & 93.90  & 88.43 & 45.57 & 51.67 & 51.01 & 53.38 & \textcolor{blue}{\textbf{45.09}} & \textcolor{red}{\textbf{42.93}} \\
    Viaduct & 109.83 & 66.8  & 68.89 & 66.65 & 36.11 & 34.32 & 37.66 & 46.14 & \textcolor{blue}{\textbf{33.55}} & \textcolor{red}{\textbf{32.81}} \\
\midrule
    Average & 126.53 & 90.40  & 93.39 & 90.86 & 53.36 & 52.55 & 56.40  & 65.51 & \textcolor{blue}{\textbf{51.08}} & \textcolor{red}{\textbf{49.45}} \\
\bottomrule[1.5pt]
    \end{tabular}%
}
  \label{aid-fid}%
\end{table*}%

\subsection{Implementation Details}
This study focus on $\times4$ SR, \emph{i.e.,} $r=4$. In our final EDiffSR, we incorporate 5 RCAB in the CPEM for prior enhancement while maintaining a favorable model size. The inner-channel number in EANet is set to $C=64$. Following prior works \cite{whl}\cite{irsde}, the depth of the noise prediction network is set to 4. In particular, the number of EAB in each depth $[{m_1},{m_2},{m_3},{m_4}]$ and $[{n_1},{n_2},{n_3},{n_4}]$ are set to $[14,1,1,1]$ and $[1,1,1,1]$, respectively. We include one EAB at the middle layer of EANet, \emph{i.e.,} $k=1$. To train our EDiffSR, we perform 500,000 iterations with a mini-batch size of 4. The initial learning rate is set to $4\times10^{-5}$ and decays following a cosine schedule. We utilize the AdamW optimizer with $\beta_1=0.9$ and $\beta_2=0.999$. The total step of the diffusion process is $T=100$. All SR methods involved in this paper were retrained from scratch on the AID training set. For a fair comparison, we did not perform any pre-training and fine-tuning processes in our EDiffSR. Our experiments are implemented on PyTorch with a 24GB memory NVIDIA RTX 3090 GPU and a 3.40 GHz AMD Ryzen 5700X CPU.

\subsection{Metrics}
In this paper, 7 metrics are used to comprehensively evaluate the performance of SR model. In the simulation experiments, where the \textcolor{blue}{ground-truth} image is available, we utilize 5 full-reference metrics: FID (Fréchet Inception Distance) \cite{fid}, LPIPS (Learned Perceptual Image Patch Similarity) \cite{lpips}, DISTS (Deep Image Structure and Texture Similarity) \cite{dists}, as well as the widely used PSNR (Peak Signal-to-Noise Ratio) and the SSIM (Structural Similarity Index) \cite{ssim}. These metrics help assess the distance between the generated images and the ground truth images. Among them, FID is widely used to measure the generative quality of the generative model. It enhances the Inception Score (IS) \cite{salimans2016improved} metric by directly measuring the feature-level distance without the need for a classifier. In real-world experiments without ground-truth images, we additionally report the results on two reference-free metrics: NIQE (Natural Image Quality Evaluator) \cite{niqe} and AG (Average Gradient). These metrics offer insights into the perceptual quality and the high-frequency details of the generated images.


\begin{figure*}[t]
\centering
\includegraphics[width=6.8in]{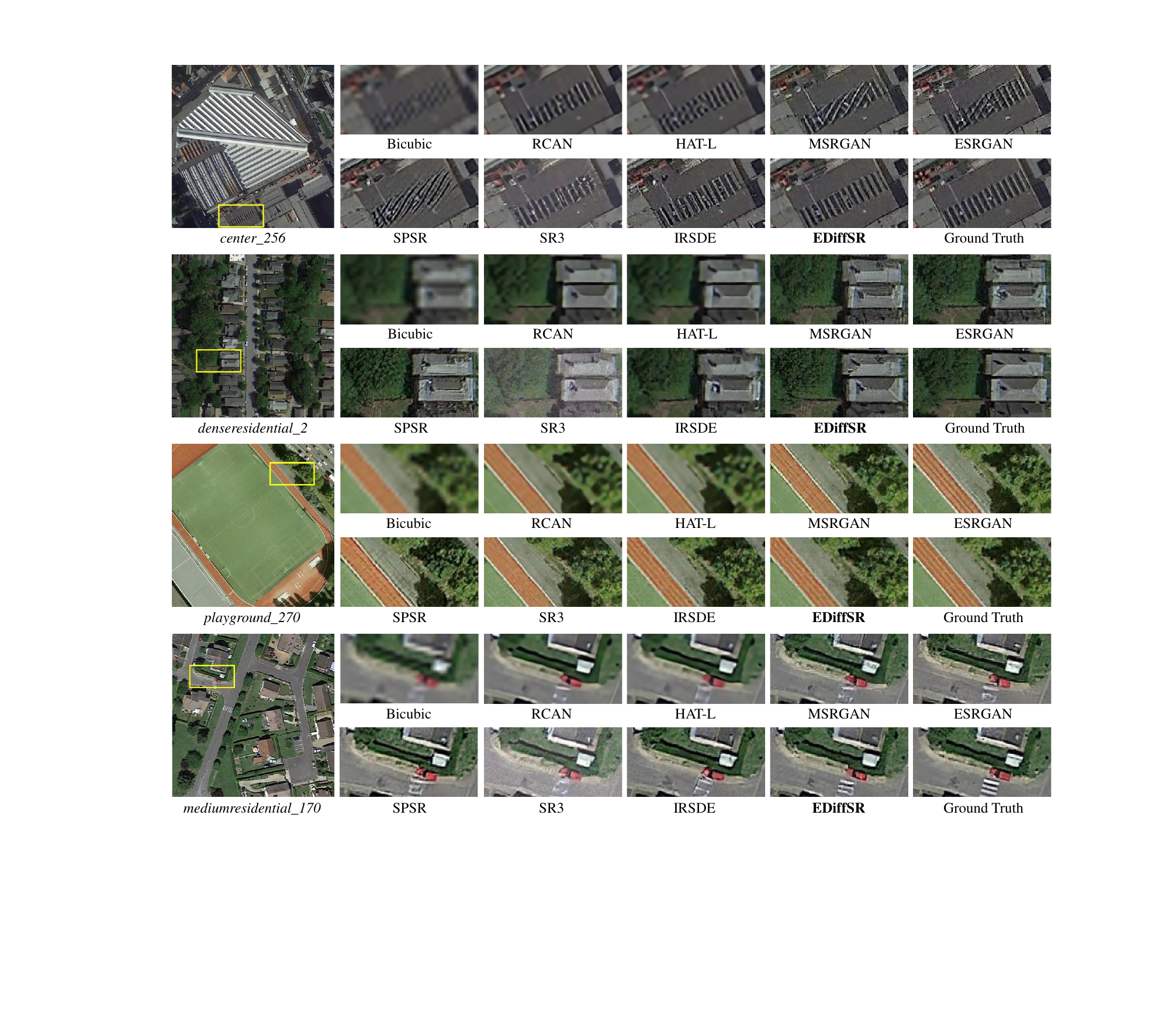}%
\captionsetup{font={scriptsize}}   
\caption{$\times4$ visual comparisons with state-of-the-art SR models on AID test set. The results show that our EDiffSR significantly outperforms comparative approaches in high-frequency detail recovery while producing visually pleasing images that are more natural. Zoom in for a better view.}
\label{vis-aid}
\end{figure*}

\begin{figure*}[t]
\centering
\includegraphics[width=7in]{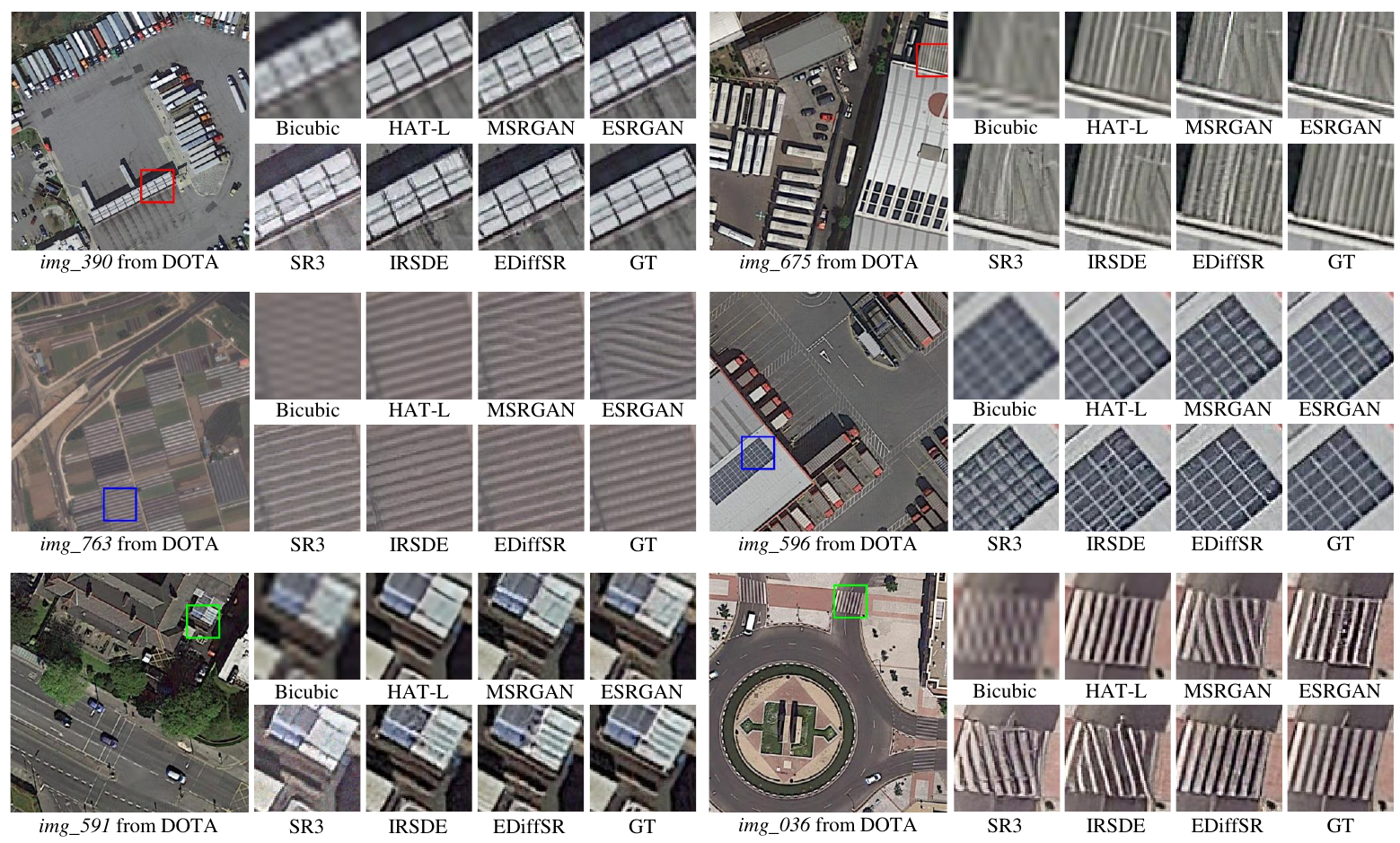}%
\captionsetup{font={scriptsize}}   
\caption{$\times4$ visual comparisons with state-of-the-art SR models on DOTA test set. Zoom in for a better view.}
\label{vis-dota}
\end{figure*}

\begin{figure*}[t]
\centering
\includegraphics[width=7in]{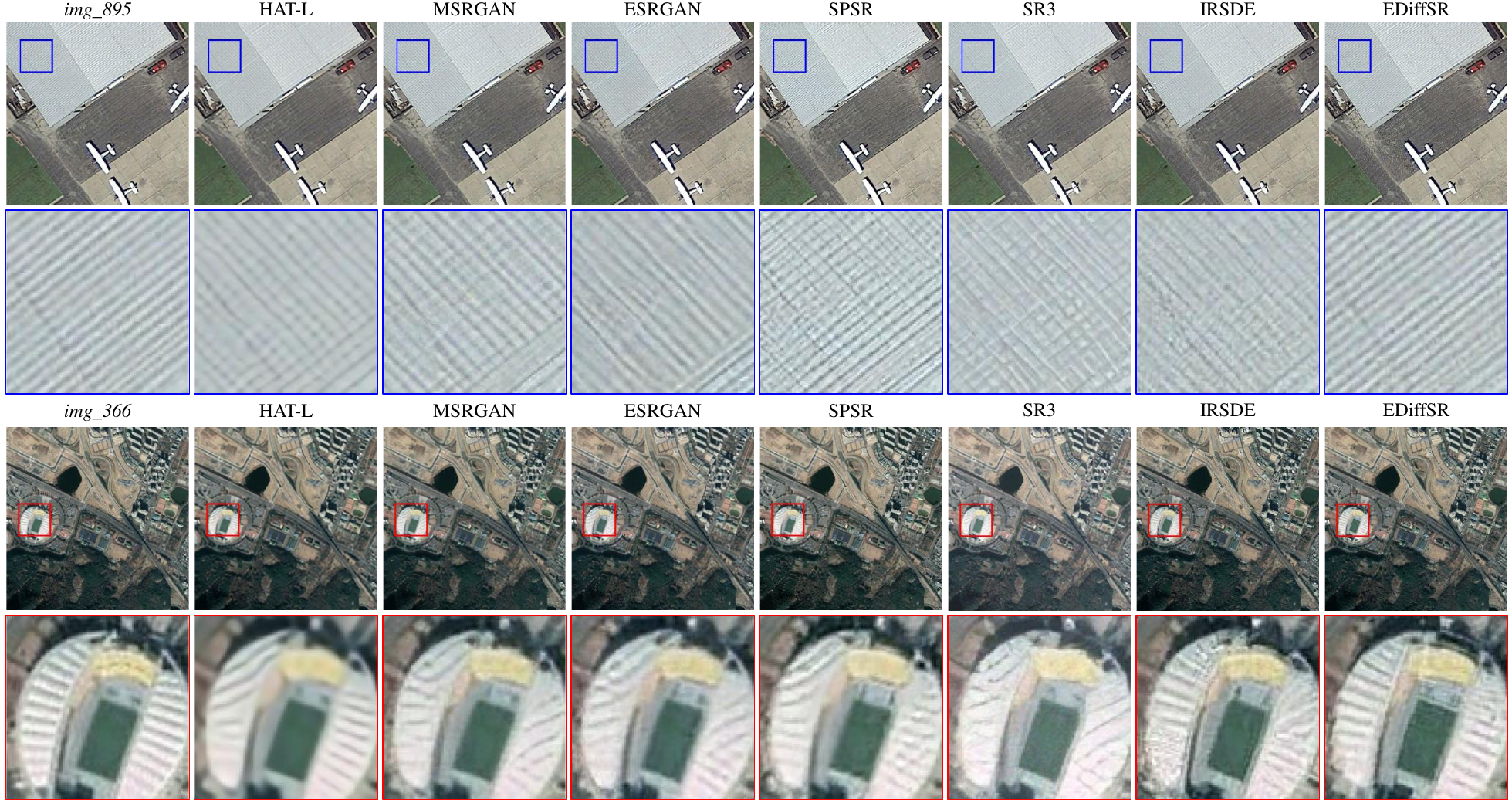}%
\captionsetup{font={scriptsize}}   
\caption{$\times4$ visual comparisons with state-of-the-art SR models on DIOR test set. Zoom in for a better view.}
\label{vis-dior}
\end{figure*}

\begin{figure}[t]
\centering
\includegraphics[width=3.3in]{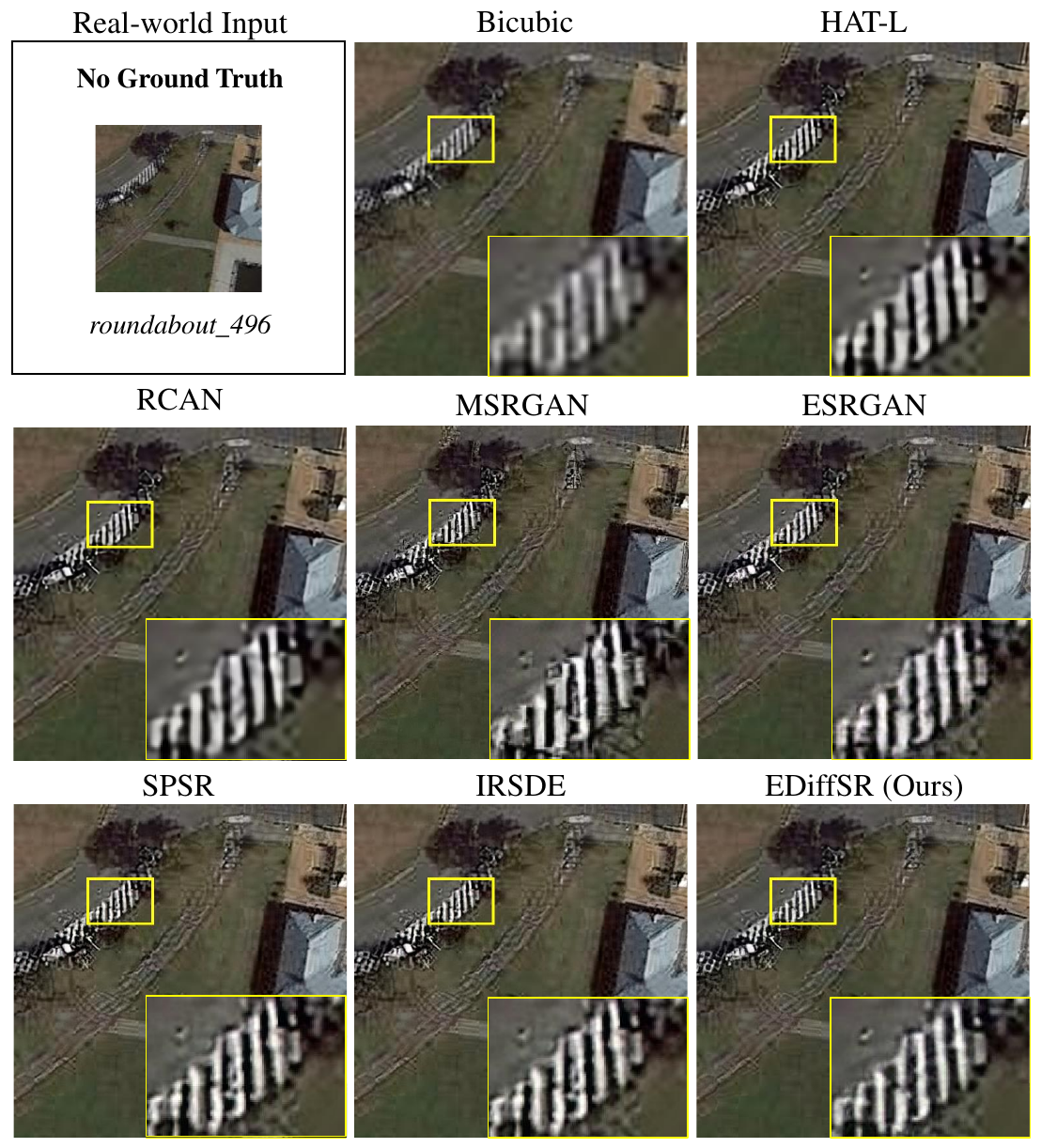}%
\captionsetup{font={scriptsize}}   
\caption{$\times4$ visual comparisons with state-of-the-art SR models on NWPU-RESISC45 with real-world degradations. Zoom in for a better view.}
\label{vis-real}
\end{figure}

\begin{figure}[t]
\centering
\includegraphics[width=3.3in]{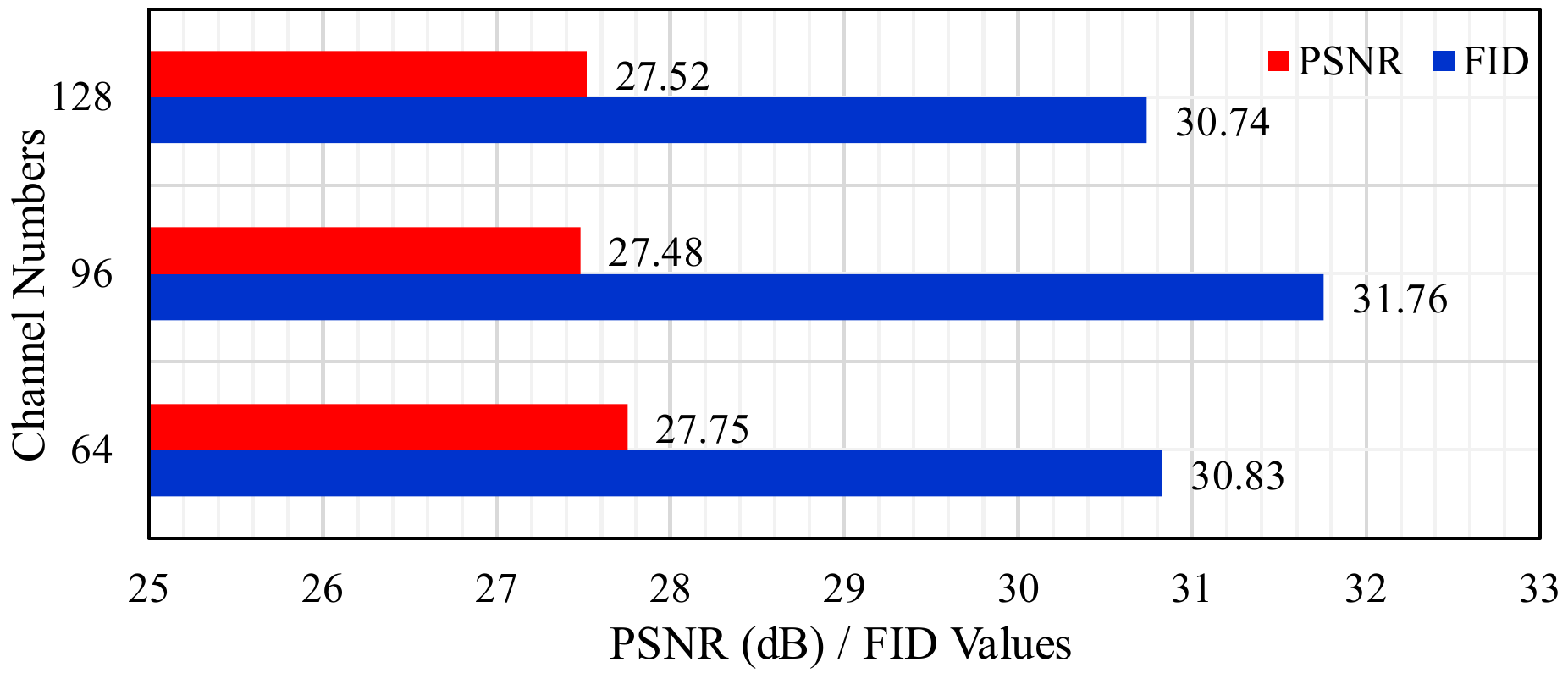}%
\captionsetup{font={scriptsize}}   
\caption{Ablation analysis of EANet with different channel numbers $C$.}
\label{channel}
\end{figure}

\begin{figure}[t]
\centering
\includegraphics[width=3.3in]{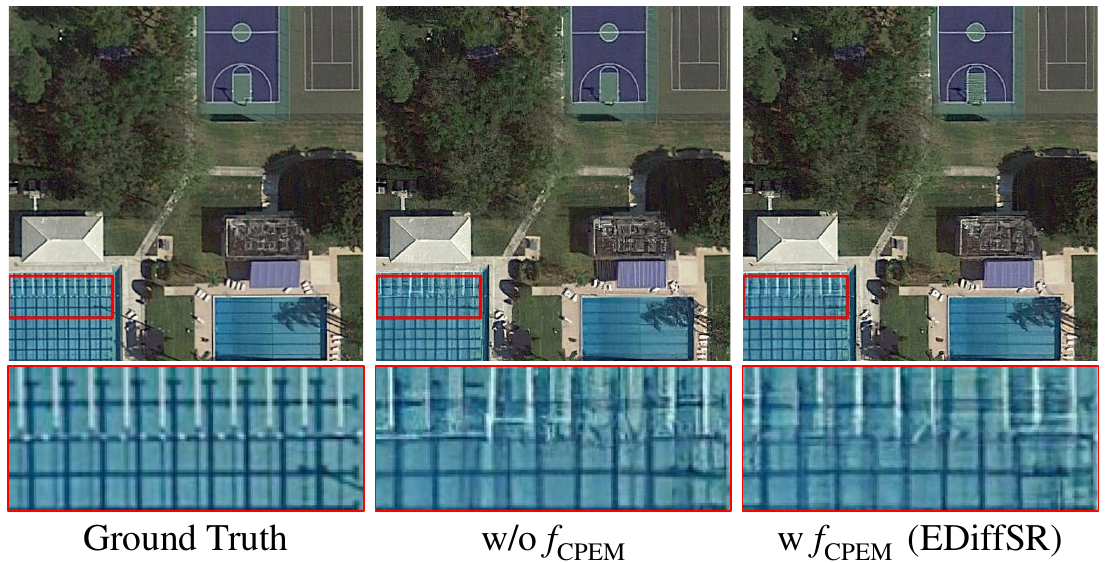}%
\captionsetup{font={scriptsize}}   
\caption{$\times4$ visual comparisons of the EDiffSR model without and with the conditional prior enhancement module $f_{CPEM}$ on "img\_074" from the DOTA test set. Image restored by the complete EDiffSR shows more high-frequency details than those recovered without $f_{CPEM}$.}
\label{aba-cpem}
\end{figure}

\subsection{Comparison With State-of-the-Arts}
We compared our EDiffSR with state-of-the-art (SOTA) SR approaches, including EDSR \cite{edsr}, RCAN \cite{rcan}, HAT-L \cite{hat}, MSRGAN \cite{srgan}, ESRGAN \cite{esrgan}, SPSR \cite{spsr}, SR3 \cite{sr3}, IRSDE \cite{irsde}. We selected these methods as they represent the mainstream approaches in the field, ensuring a comprehensive evaluation. Specifically, EDSR, RCAN, and HAT-L are CNN-based approaches that adopt wide CNN, channel attention, and transformer architectures, respectively. Note that MSRGAN is a modified version of SRGAN, where the BN layer is removed to avoid artifacts, and it employs the same perceptual loss as ESRGAN. SPSR employs carefully designed gradient loss to preserve structural details and has demonstrated favorable performance. On the other hand, SR3 and IRSDE are SOTA diffusion-based models. We retrained these comparative approaches on the AID training set according to their official implementation settings.

\subsubsection{Quantitative Comparison}
Results of FID values on 30 categories of the AID test set are reported in Table. \ref{aid-fid}. In each row, we highlighted the best and the second-best FID performance. We can find in most remote sensing scenes our EDiffSR achieves favorable FID performance against all comparative models. However, due to the complex diversity of remote sensing scenes, achieving generalization across various scenes remains a challenging task. Specifically, EDiffSR outperforms the second-best approach (IRSDE) by an average margin of 1.63 in terms of FID. These results reveal that EDiffSR can provide robust high-quality data distribution in various remote sensing scenarios, highlighting its favorable generative capability.

Additionally, we presented the average FID, LPIPS, DISTS, and NIQE values across the AID, DOTA, and DIOR test sets in Table \ref{res-aid-dota-dior}. We observed that our EDiffSR still achieves the best FID performance across these test sets. It is worth highlighting that GAN-based models excel in achieving the best LPIPS results because they usually adopt the VGG space to compute the perceptual loss, which aligns with the calculation of LPIPS. In this case, our EDiffSR achieves acceptable LPIPS results and surpasses Diffusion-based approaches by a large margin. For instance, compared to SR3, EDiffSR exhibits a remarkable 0.0647 improvement in terms of LPIPS. When compared to IRSDE, we achieve superior LPIPS performance (0.1898 vs. 0.2419) in the DIOR dataset. Notably, both IRSDE and EDiffSR utilize the same diffusion process equation, \emph{i.e.,} SDE. Therefore, the results demonstrate the superiority of our EANet in providing effective noise prediction capabilities compared to the commonly employed UNet architecture in IRSDE. As for the DISTS metrics, we observe that despite SPSR focusing on preserving structural details, it only achieves the second-best performance on the DOTA and DIOR datasets. In contrast, our EDiffSR excels in attaining the best DISTS scores for both DOTA and DIOR, highlighting its remarkable capability to restore accurate structural details in remote sensing images. Moreover, EDiffSR can achieve the best NIQE values in almost all test sets. As a result, the proposed EDiffSR does recover realistic results that align well with human perception.

Besides the above results, the PSNR and SSIM results are also tabulated in Table. \ref{res-psnr}. The best and second-best performance within each category of methods are highlighted in bold and underlined. PSNR-oriented models such as EDSR, RCAN, and HAT-L achieve higher PSNR and SSIM scores when compared to GAN-based and DPM-based SR methods. This is because they optimize Mean Squared Error (MSE), which provides a straightforward learning objective for high PSNR performance. In particular, PSNR and SSIM are inconsistent with human perceptual while guiding the network to generate over-smooth content. As demonstrated in Table. \ref{res-aid-dota-dior}, despite HAT-L achieving the highest PSNR score, it gains undesired scores in terms of FID, DISTS, and LPIPS. Furthermore, PSNR-driven methods tend to produce blurry results, resulting in poor perceptual quality. In the DPM-based category, our EDiffSR consistently delivers higher-quality results while maintaining the best PSNR/SSIM performance. When compared to SR3, we achieve a significant improvement in terms of PSNR (27.40dB vs. 26.24dB) in the AID test set, demonstrating that our lightweight EANet is capable of providing excellent denoising performance in SR tasks.

\begin{table*}[t]
  \centering
\captionsetup{font={scriptsize}}   
  \caption{Quantitative comparison with state-of-the-art SR models in terms of FID, LPIPS, DISTS, and NIQE across AID, DOTA, and DIOR test sets. The best performance value is highlighted in \textcolor{red}{\textbf{red}} while the second best is in \textcolor{blue}{\textbf{blue}}.}
\setlength{\tabcolsep}{1.3mm}{
    \begin{tabular}{cccccccccccc}
\toprule[1.5pt]
    \multirow{2}[2]{*}{Dataset} & \multirow{2}[2]{*}{Metrics} & \multicolumn{1}{c}{Baseline}& \multicolumn{3}{c}{CNN-based} & \multicolumn{3}{c}{GAN-based} & \multicolumn{3}{c}{Diffusion-based} \\
 \cmidrule{3-12}      &       & Bicubic & EDSR \cite{edsr} & RCAN \cite{rcan} & \textcolor{blue}{HAT-L} \cite{hat}   & MSRGAN \cite{esrgan} & ESRGAN \cite{esrgan} & SPSR \cite{spsr} & SR3 \cite{sr3}  & IRSDE \cite{irsde} & EDiffSR \\
\midrule \midrule
    \multirow{4}[0]{*}{AID \cite{aid}} & FID $\downarrow$   & 126.53  &90.40 & 93.39  & 90.86 & 56.36  & 52.55  & 56.40  & 65.51  & \textcolor{blue}{\textbf{51.08}} & \textcolor{red}{\textbf{49.45}}  \\
          & LPIPS $\downarrow$& 0.4801  &0.3068 & 0.3112 & 0.3078 & \textcolor{red}{\textbf{0.1694}}  & \textcolor{blue}{\textbf{0.1695}} & 0.1751 & 0.2534 & 0.2247 & 0.1887 \\
          & DISTS $\downarrow$& 0.1512  & 0.0880& 0.0900  & 0.0882 & 0.0590  & 0.0601  & 0.0632  & 0.0903  & \textcolor{blue}{\textbf{0.0579}} & \textcolor{red}{\textbf{0.0561}}  \\
          & NIQE $\downarrow$ & 21.24  &19.79 & 19.88  & 20.15 & 17.46  & 15.51  & 17.90  & 15.16  & \textcolor{blue}{\textbf{14.80}} & \textcolor{red}{\textbf{14.22}}  \\
\midrule \midrule
    \multirow{4}[0]{*}{DOTA \cite{edsr}} & FID $\downarrow$   & 65.99 &55.22 & 43.37 & 42.61 & 25.12 & 24.42 & 26.42 &   35.74    & \textcolor{blue}{\textbf{23.91}} & \textcolor{red}{\textbf{21.26}} \\
          & LPIPS $\downarrow$& 0.4416 & 0.2616& 0.2629 & 0.2641 & 0.1649 & \textcolor{red}{\textbf{0.1506}} & \textcolor{blue}{\textbf{0.1605}} &    0.2790   & 0.1919 & 0.1689 \\
          & DISTS $\downarrow$& 0.1481 & 0.0831& 0.0846 & 0.0832 & 0.0589 & 0.0582 & 0.0617 &    0.0993   & \textcolor{red}{\textbf{0.0550}} &  \textcolor{blue}{\textbf{0.0556}}\\
          & NIQE  $\downarrow$& 19.48 & 18.37& 18.58 & 18.98 & 16.71 & 15.48 & 18.01 &   15.49    & \textcolor{blue}{\textbf{14.47}} & \textcolor{red}{\textbf{14.20}} \\
\midrule \midrule
    \multirow{4}[0]{*}{DIOR \cite{dior}} & FID $\downarrow$  & 57.42 &41.87 & 43.20  & 41.94 & 23.16 & 22.51 & 24.02 &   29.60    & \textcolor{blue}{\textbf{22.28}} & \textcolor{red}{\textbf{21.79}} \\
          & LPIPS $\downarrow$& 0.4678 & 0.3020& 0.3048 & 0.3062 & \textcolor{red}{\textbf{0.1722}} & 0.1836 & \textcolor{blue}{\textbf{0.1772}} &   0.2836    & 0.2419 & 0.1898 \\
          & DISTS $\downarrow$& 0.1497 & 0.0886& 0.0899 & 0.0893 & \textcolor{blue}{\textbf{0.0605}} & 0.0623 & 0.0654 &    0.0924   & 0.0622 &  \textcolor{red}{\textbf{0.0590}} \\
          & NIQE $\downarrow$ & 20.24 &19.16 & 19.30  & 19.56 & 17.61 & \textcolor{red}{\textbf{15.13}} & 18.24 &    15.55   & 15.19 & \textcolor{blue}{\textbf{15.16}} \\
\bottomrule[1.5pt]
    \end{tabular}%
}
  \label{res-aid-dota-dior}%
\end{table*}%

\begin{table*}[t]
  \centering
\captionsetup{font={scriptsize}}   
  \caption{Quantitative comparison with state-of-the-art SR models in terms of NIQE, and AG on NWPU-RESISC45 test set. The best performance value is highlighted in \textcolor{red}{\textbf{red}} while the second best in \textcolor{blue}{\textbf{blue}}.}
\setlength{\tabcolsep}{2.8mm}{
    \begin{tabular}{cccccccccc}
\toprule[1.5pt]
    Metrics & Bicubic & RCAN \cite{rcan}  & HAT-L \cite{hat} & MSRGAN \cite{srgan} & ESRGAN \cite{esrgan} & SPSR \cite{spsr}  & SR3 \cite{sr3}   & IRSDE \cite{irsde} & EDiffSR \\
\midrule
    NIQE $\downarrow$  & 20.8032 & 20.4722 & 20.3153 & 17.3114 & 17.9991 & 18.0730 & 17.8176 & \textcolor{blue}{\textbf{16.5357}} & \textcolor{red}{\textbf{16.3013}} \\
    AG $\uparrow$    & 2.3556 & 3.0296 & 3.0081 & 3.3771 & 3.5170 & 3.5723 & 4.0101 & \textcolor{blue}{\textbf{4.2690}} & \textcolor{red}{\textbf{4.7461}} \\
\bottomrule[1.5pt]
    \end{tabular}%
}
  \label{res-real}%
\end{table*}%

\begin{table}[!t]
  \centering
\captionsetup{font={scriptsize}}   
  \caption{Quantitative comparison with state-of-the-art SR models in terms of PSNR, and SSIM across AID, DOTA, and DIOR test sets. The best performance value in each model type is highlighted in \textbf{blod} while the second best is in \underline{underline}.}
\setlength{\tabcolsep}{1.5mm}{
    \begin{tabular}{ccccccc}
\toprule[1.5pt]
    \multirow{2}[2]{*}{Method} & \multicolumn{2}{c}{AID \cite{aid}} & \multicolumn{2}{c}{DOTA \cite{dota}} & \multicolumn{2}{c}{DIOR \cite{dior}} \\
   \cmidrule{2-7} & PSNR  & SSIM  & PSNR  & SSIM  & PSNR  & SSIM \\
\midrule
    EDSR \cite{edsr}  &    30.65   &   0.8085    &       33.64    &   0.8648       &  30.63   & 0.8116 \\
    RCAN\cite{rcan}  &    \textbf{30.82}   &    \underline{0.8121}   &   \underline{33.86}   &   \underline{0.8680}    &\underline{30.85} & \underline{0.8159}  \\
    HAT-L \cite{hat} &   \underline{30.81}    &   \textbf{0.8124}    &  \textbf{33.99}      &  \textbf{0.8684}     &  \textbf{30.87}    &\textbf{0.8161}  \\
\midrule
    MSRGAN \cite{srgan} &  \textbf{28.75}     &  \textbf{0.7390}     &   \textbf{29.46}    &   \textbf{0.7825}    &  \textbf{28.84}     &\textbf{0.7422}  \\
    ESRGAN \cite{esrgan} &    \underline{28.38}   & \underline{0.7272}      &   \underline{29.01}    & \underline{0.7716}      &  \underline{28.07}    &  \underline{0.7086} \\
    SPSR \cite{spsr} &   27.71    &    0.7081   &  28.00 & 0.7696     &  27.46 & 0.7106  \\
\midrule
    SR3 \cite{sr3}   &    26.24    &  \underline{0.6705}     &    27.62    & 0.6804      & 26.18 & \underline{0.6639}    \\
    IRSDE \cite{irsde}&   \underline{27.19}    &    0.6585   &    \underline{28.08}   &  \underline{0.7133}     &   \underline{27.11} & 0.6516     \\
    EDiffSR (Ours) &    \textbf{27.40}   &    \textbf{0.6805}   &    \textbf{28.30}     &   \textbf{0.7345}    &  \textbf{27.55} & \textbf{0.6823}    \\
\bottomrule[1.5pt]
    \end{tabular}%
}
  \label{res-psnr}%
\end{table}%

\subsubsection{Qualitative Comparison}
We conducted a visual comparison with all comparative models. From Fig. \ref{vis-aid}, we find that our EDiffSR can consistently produce photo-realistic results that surpass SOTA approaches. For the "center\_256" in AID, both RCAN and HAT-L produce blurry results, highlighting the limited generalization of PSNR-oriented models in recovering rich details. In contrast, GAN-based models can restore shape details, especially edge information, but often introduce severe artifacts inconsistent with the ground truth. EDiffSR consistently delivers more natural and realistic results, demonstrating its capacity to generate visually pleasing images. For the "mediumresidential\_170" image, RCAN and HAT-L still exhibit an over-smoothed appearance, while other GAN-based and diffusion-based models yield more natural results with realistic details. Compared with IRSDE  and SR3 which directly adopt the bicubic interpolation to prepare conditions, EdiffSR contains more context details that are close to the ground-truth image, such as the the marks on the road. The results demonstrate that the proposed CPEM is helpful in exploring more useful priors, \emph{e.g.,} edge information, thus boosting the performance of DPM in SR tasks.

In Fig. \ref{vis-dota}, we also visualize some SR results on the DOTA test set. As shown in Fig. \ref{vis-dota}, both CNN-based approaches and GAN-based models fail to achieve satisfactory detail, especially in terms of edges and textures. For "img\_591" from DOTA, only EDiffSR successfully restores the clear and sharp details of the building on the ground. In "img\_036", MSRGAN, SR3, and IRSDE exhibit severe distortion, which deviates from the ground-truth distribution. HAT-L and ESRGAN can offer relatively realistic distribution, restoring accurate direction of the lines on the road. Nevertheless, the results obtained from ESRGAN exhibit an unnatural appearance due to the oversharpening issue. In contrast, EDiffSR accurately generates these details and appears more natural perception. These results highlight the capability of CPEM to explore additional prior information, enabling EDiffSR to recover more details that align with the realistic distribution of the ground truth.

In Fig. \ref{vis-dior}, we zoomed in and displayed some visual results from the DOIR dataset. As shown in "img\_895", EDiffSR exhibits an impressive visual performance, outperforming other methods in accurately recovering the direction of the lines on the building roof. In this context, reconstructing such high-frequency information can be challenging. All methods yielded a completely wrong distribution of these details, except our EDiffSR. Benefiting from our condition prior enhancement module (CPEM), more high-frequency prior information can be explored and introduced into the diffusion process, making the output images consistent with the spatial distribution of ground-truth images. In “img\_366”, we zoomed in on the stadium region for comparison. It is evident that our EDiffSR reconstructs the most realistic results, whereas the other models exhibit significant distortion and blurring.

\begin{table}[t]
  \centering
\captionsetup{font={scriptsize}}   
  \caption{Ablation analysis of EDiffSR with different components. The best FID performance is shown in \textbf{blod}.}
\setlength{\tabcolsep}{2mm}{
    \begin{tabular}{cccccc}
\toprule[1.5pt]
    Methods & $f_{CPEM}$ & EANet   & UNet  & Param. (M) & FID $\downarrow$ \\
\midrule
    Baseline &   	\XSolidBrush   &   \XSolidBrush   &     \Checkmark  & 137.15 & 32.42  \\
    Model-1 &   \Checkmark  &   \XSolidBrush    &     \Checkmark  & 137.59& 32.68 \\
    Model-2 &   \XSolidBrush  &   \Checkmark    &    \XSolidBrush   &  26.34& 31.11\\
    EDiffSR (Ours)  &    \Checkmark   &    \Checkmark   &   \XSolidBrush    &26.79 & \textbf{30.83}\\
\bottomrule[1.5pt]
    \end{tabular}%
}
  \label{component}%
\end{table}%

\begin{table}[t]
  \centering
\captionsetup{font={scriptsize}}   
  \caption{Ablation analysis of EANet with different scales of convolution. $3\times3$, $5\times5$, and $7\times7$ represents single-scale design with different DWConv kernels. EDiffSR adopts the multi-scale design and achieves modest improvement.}
\setlength{\tabcolsep}{2.5mm}{
    \begin{tabular}{ccccc}
\toprule[1.5pt]
      Methods    & $3\times3$     & $5\times5$     & $7\times7$     & EDiffSR (Ours) \\
\midrule
    FID $\downarrow$   &   30.98    &  31.17     &   31.06    &\textbf{30.83}  \\
    PSNR $\uparrow$ &   27.59    &  27.77     & \textbf{27.89}      &27.75  \\
\bottomrule[1.5pt]
    \end{tabular}%
}
  \label{multiscale}%
\end{table}%

\begin{table}[t]
  \centering
\captionsetup{font={scriptsize}}   
  \caption{Model efficiency analysis with state-of-the-art SR models. The best performance is shown in \textbf{blod}.}
\setlength{\tabcolsep}{1.5mm}{
    \begin{tabular}{ccccc}
\toprule[1.5pt]
    Methods & Param. (M)  & Running Time (s)  & FID $\downarrow$ & PSNR (dB) $\uparrow$  \\
\midrule
    EDSR \cite{edsr}   &   43.09    &   0.93    &   62.50    & 31.64  \\
    RCAN \cite{rcan}  &   15.59    &     0.29  &    59.99   & 31.84  \\
    HAT-L \cite{hat} &    40.32   &    0.76   &    58.47   &  \textbf{31.89} \\
\midrule
    MSRGAN \cite{srgan} &   \textbf{1.52}    &   \textbf{0.19}    &  34.88     &29.02   \\
    ESRGAN \cite{esrgan} &    16.70   &   0.22    &  33.16    & 28.49 \\
    SPSR \cite{spsr} &   24.79    &   0.60    &     35.61  & 27.72 \\
\midrule
    SR3 \cite{sr3}  &   92.56    &    137.61   &    43.61   &  26.68\\
    IRSDE \cite{irsde} &    137.15   &   27.90    &   32.42    &  27.46\\
    EDiffSR (Ours) & 26.79 &  19.26   &     \textbf{30.83}  &   27.75      \\
\bottomrule[1.5pt]
    \end{tabular}%
}
  \label{efficiency}%
\end{table}%

\subsubsection{Real-world Comparison}
We also evaluate the performance of our EDiffSR on real-world remote sensing images, \emph{i.e.,} without performing simulated degradations. Table. \ref{res-real} shows the quantitative comparison of EDiffSR against SOTA methods in terms of NIQE and AG. We can see that EDiffSR achieves the best NIQE performance, illustrating our method can restore natural images that align with human perception in real-world scenarios. In addition, the best AG performance demonstrates that our reconstructed image contains more high-frequency detail information, such as edges and textures.

More intuitively, we display the visual comparison on the NWPU-RESISC45 dataset. The qualitative results are shown in Fig. \ref{vis-real}. We can see that the PSNR-driven approach HAT-L exhibits a significantly blurry effect compared to the other approaches. Whereas GAN-based methods such as MSRGAN produce excessively sharpened results accompanied by pseudo-details. In the diffusion-based methods, SR3 shows limitations in recovering precise edge information of the dense lines on the ground. In contrast, our method demonstrates the clearest preservation of high-frequency texture information, with minimal blurring and artifacts.

\subsection{Ablation Studies}
In this section, we conduct extensive experiments to demonstrate the effectiveness of each component within our EDiffSR. Note that the FID and PSNR values are the average results of the AID, DOTA, and DIOR datasets.
\subsubsection{Component Analysis of EDiffSR}
To investigate the holistic effectiveness of each part within EDiffSR, we remove the conditional prior enhancement ($f_{CPEM}$), the Efficient Activate Network (EANet) to form the three models reported in Table. \ref{component}. Note that once we remove the EANet, we replace it with the Vanillia UNet for noise prediction. By comparing Model-1 and EDiffSR, we can find that EANet is superior in improving the FID performance than UNet (30.83 vs. 32.68) while reducing the model size by a large margin (26.31M vs. 137.15M). After adding the  $f_{CPEM}$ in Model-2, we observe a slight parameter increase, but the improvement in FID is significant. When the whole $f_{CPEM}$ and EANet are absent in Baseline, the model performs poorly in FID. These results demonstrate that the proposed $f_{CPEM}$ and EANet are able to improve the performance of the diffusion model. Besides, both $f_{CPEM}$ and EANet have low complexity, allowing EDiffSR efficient yet effective.

\subsubsection{Effectiveness of EANet}
We first investigate the impact of varying channel numbers of EANet. As shown in Fig. \ref{channel}, we find that EDiffSR achieves slightly superior FID performance when $C=128$ compared to $C=64$. However, it was observed that EDiffSR yields the highest PSNR results when $C=64$. To strike a favorable balance between model size and performance, we set $C=64$ in our final EDiffSR. Moreover, we conducted three experiments to assess the impact of multi-scale design in EAB. The results are listed in Table. \ref{multiscale}. From the table, we can see that the multi-scale design brings a modest improvement in terms of FID.

\subsubsection{Effectiveness of Conditional Prior Enhancement}
To further illustrate the capability of $f_{CPEM}$ in grasping valuable prior knowledge for accurate SR reconstruction, we provide a visual comparison in Figure \ref{aba-cpem}. From this figure, we can see that the model with $f_{CPEM}$ excels in recovering high-frequency details, such as edges and boundaries. This observation highlights that $f_{CPEM}$ indeed boosts the performance of EDiffSR by exploring an enriched condition with more priors.

\subsubsection{Model Efficiency}
To demonstrate the efficiency of EDiffSR, we conducted a comparison of parameters and inference times, as presented in Table \ref{efficiency}. The results indicate that EDiffSR is far more lightweight compared to existing DPM-based SR models. For instance, when compared with IRSDE, EDiffSR achieves an impressive reduction of nearly 80\% in model parameters (26.79M vs. 137.15M) while delivering superior performance in both FID (30.83 vs. 32.42) and PSNR (27.75dB vs. 27.46dB). Furthermore, EDiffSR exhibits faster inference compared to existing DPM-based models. When compared to SR3, EDiffSR is 7 times faster (19.26s vs. 137.61s) in the diffusion sampling process, making it more practical for real-world applications.

\section{Conclusion}\label{conclu}
In this paper, we devise an efficient diffusion probabilistic model (EDiffSR) to generate perceptual-pleasant SR results of remote sensing images. The proposed efficient activation network (EANet) shows superior performance against vanilla UNet in noise prediction and is more lightweight. In particular, rather than employing the interpolated condition, a condition prior enhancement module is designed to explore the potential priors from LR input, which significantly boosts the reconstruction performance. Rigorous quantitative and qualitative evaluation on AID, DOTA, DIOR, and NWPU-RESISC45 datasets demonstrated our EDiffSR outperforms state-of-the-art CNN-based, GAN-based, and Diffusion-based SR methods in data distribution and perceptual quality.

Nevertheless, our EDiffSR does exhibit some shortcomings. Firstly, the sampling process of the diffusion model consumes massive computational costs, which hinders its real-time application. Secondly, EDiffSR does not consider the multiple degradations involved in remote sensing images, resulting in limited adaptability to real-world scenes. Therefore, more efforts should be paid to speed up the sampling process of the diffusion model in the future direction. Moreover, we consider extending our EDiffSR to blind SR issues, thus improving its generalization in real-world scenarios.



%
%



%

%


%
%

\ifCLASSOPTIONcaptionsoff
  \newpage
\fi



%
\bibliographystyle{IEEEtran}
\bibliography{reference}

\begin{thebibliography}{10}
\providecommand{\url}[1]{#1}
\csname url@samestyle\endcsname
\providecommand{\newblock}{\relax}
\providecommand{\bibinfo}[2]{#2}
\providecommand{\BIBentrySTDinterwordspacing}{\spaceskip=0pt\relax}
\providecommand{\BIBentryALTinterwordstretchfactor}{4}
\providecommand{\BIBentryALTinterwordspacing}{\spaceskip=\fontdimen2\font plus
\BIBentryALTinterwordstretchfactor\fontdimen3\font minus \fontdimen4\font\relax}
\providecommand{\BIBforeignlanguage}[2]{{%
\expandafter\ifx\csname l@#1\endcsname\relax
\typeout{** WARNING: IEEEtran.bst: No hyphenation pattern has been}%
\typeout{** loaded for the language `#1'. Using the pattern for}%
\typeout{** the default language instead.}%
\else
\language=\csname l@#1\endcsname
\fi
#2}}
\providecommand{\BIBdecl}{\relax}
\BIBdecl

\bibitem{fid}
M.~Heusel, H.~Ramsauer, T.~Unterthiner, B.~Nessler, and S.~Hochreiter, ``Gans trained by a two time-scale update rule converge to a local nash equilibrium,'' \emph{Advances in neural information processing systems}, vol.~30, 2017.

\bibitem{edsr}
B.~Lim, S.~Son, H.~Kim, S.~Nah, and K.~Mu~Lee, ``Enhanced deep residual networks for single image super-resolution,'' in \emph{Proceedings of the IEEE conference on computer vision and pattern recognition workshops}, 2017, pp. 136--144.

\bibitem{rcan}
Y.~Zhang, K.~Li, K.~Li, L.~Wang, B.~Zhong, and Y.~Fu, ``Image super-resolution using very deep residual channel attention networks,'' in \emph{Proceedings of the European conference on computer vision (ECCV)}, 2018, pp. 286--301.

\bibitem{hat}
X.~Chen, X.~Wang, J.~Zhou, Y.~Qiao, and C.~Dong, ``Activating more pixels in image super-resolution transformer,'' in \emph{Proceedings of the IEEE/CVF Conference on Computer Vision and Pattern Recognition}, 2023, pp. 22\,367--22\,377.

\bibitem{srgan}
C.~Ledig, L.~Theis, F.~Husz{\'a}r, J.~Caballero, A.~Cunningham, A.~Acosta, A.~Aitken, A.~Tejani, J.~Totz, Z.~Wang \emph{et~al.}, ``Photo-realistic single image super-resolution using a generative adversarial network,'' in \emph{Proceedings of the IEEE conference on computer vision and pattern recognition}, 2017, pp. 4681--4690.

\bibitem{esrgan}
X.~Wang, K.~Yu, S.~Wu, J.~Gu, Y.~Liu, C.~Dong, Y.~Qiao, and C.~Change~Loy, ``Esrgan: Enhanced super-resolution generative adversarial networks,'' in \emph{Proceedings of the European conference on computer vision (ECCV) workshops}, 2018, pp. 1--17.

\bibitem{spsr}
C.~Ma, Y.~Rao, J.~Lu, and J.~Zhou, ``Structure-preserving image super-resolution,'' \emph{IEEE transactions on pattern analysis and machine intelligence}, vol.~44, no.~11, pp. 7898--7911, 2021.

\bibitem{sr3}
C.~Saharia, J.~Ho, W.~Chan, T.~Salimans, D.~J. Fleet, and M.~Norouzi, ``Image super-resolution via iterative refinement,'' \emph{IEEE Transactions on Pattern Analysis and Machine Intelligence}, vol.~45, no.~4, pp. 4713--4726, 2023.

\bibitem{irsde}
Z.~Luo, F.~K. Gustafsson, Z.~Zhao, J.~Sj{\"o}lund, and T.~B. Sch{\"o}n, ``Image restoration with mean-reverting stochastic differential equations,'' in \emph{International Conference on Machine Learning}.\hskip 1em plus 0.5em minus 0.4em\relax PMLR, 2023.

\bibitem{he2023spectral}
J.~He, Q.~Yuan, J.~Li, Y.~Xiao, D.~Liu, H.~Shen, and L.~Zhang, ``Spectral super-resolution meets deep learning: Achievements and challenges,'' \emph{Information Fusion}, p. 101812, 2023.

\bibitem{bsvsr}
Y.~Xiao, Q.~Yuan, Q.~Zhang, and L.~Zhang, ``Deep blind super-resolution for satellite video,'' \emph{IEEE Transactions on Geoscience and Remote Sensing}, vol.~61, pp. 1--16, 2023.

\bibitem{liu2023efficient}
D.~Liu, J.~Li, Q.~Yuan, L.~Zheng, J.~He, S.~Zhao, and Y.~Xiao, ``An efficient unfolding network with disentangled spatial-spectral representation for hyperspectral image super-resolution,'' \emph{Information Fusion}, vol.~94, pp. 92--111, 2023.

\bibitem{drw1}
R.~Dian, A.~Guo, and S.~Li, ``Zero-shot hyperspectral sharpening,'' \emph{IEEE Transactions on Pattern Analysis and Machine Intelligence}, vol.~45, no.~10, pp. 12\,650--12\,666, 2023.

\bibitem{hd1}
D.~He and Y.~Zhong, ``Deep hierarchical pyramid network with high-frequency-aware differential architecture for super-resolution mapping,'' \emph{IEEE Transactions on Geoscience and Remote Sensing}, vol.~61, pp. 1--15, 2023.

\bibitem{hd2}
D.~He, Q.~Shi, X.~Liu, Y.~Zhong, G.~Xia, and L.~Zhang, ``Generating annual high resolution land cover products for 28 metropolises in china based on a deep super-resolution mapping network using landsat imagery,'' \emph{GIScience \& Remote Sensing}, vol.~59, no.~1, pp. 2036--2067, 2022.

\bibitem{he2023self}
J.~He, Q.~Yuan, J.~Li, Y.~Xiao, and L.~Zhang, ``A self-supervised remote sensing image fusion framework with dual-stage self-learning and spectral super-resolution injection,'' \emph{ISPRS Journal of Photogrammetry and Remote Sensing}, vol. 204, pp. 131--144, 2023.

\bibitem{dian2020regularizing}
R.~Dian, S.~Li, and X.~Kang, ``Regularizing hyperspectral and multispectral image fusion by cnn denoiser,'' \emph{IEEE transactions on neural networks and learning systems}, vol.~32, no.~3, pp. 1124--1135, 2020.

\bibitem{xiao2022generating}
Y.~Xiao, Y.~Wang, Q.~Yuan, J.~He, and L.~Zhang, ``Generating a long-term (2003- 2020) hourly 0.25° global pm2. 5 dataset via spatiotemporal downscaling of cams with deep learning (deepcams),'' \emph{Science of The Total Environment}, vol. 848, p. 157747, 2022.

\bibitem{xiao2022space}
Y.~Xiao, Q.~Yuan, J.~He, Q.~Zhang, J.~Sun, X.~Su, J.~Wu, and L.~Zhang, ``Space-time super-resolution for satellite video: A joint framework based on multi-scale spatial-temporal transformer,'' \emph{International Journal of Applied Earth Observation and Geoinformation}, vol. 108, p. 102731, 2022.

\bibitem{10144690}
Q.~Zhang, Y.~Zheng, Q.~Yuan, M.~Song, H.~Yu, and Y.~Xiao, ``Hyperspectral image denoising: From model-driven, data-driven, to model-data-driven,'' \emph{IEEE Transactions on Neural Networks and Learning Systems}, pp. 1--21, 2023.

\bibitem{miao2021hyperspectral}
Y.-C. Miao, X.-L. Zhao, X.~Fu, J.-L. Wang, and Y.-B. Zheng, ``Hyperspectral denoising using unsupervised disentangled spatiospectral deep priors,'' \emph{IEEE Transactions on Geoscience and Remote Sensing}, vol.~60, pp. 1--16, 2021.

\bibitem{jiang2020hierarchical}
K.~Jiang, Z.~Wang, P.~Yi, and J.~Jiang, ``Hierarchical dense recursive network for image super-resolution,'' \emph{Pattern Recognition}, vol. 107, p. 107475, 2020.

\bibitem{rdn}
Y.~Zhang, Y.~Tian, Y.~Kong, B.~Zhong, and Y.~Fu, ``Residual dense network for image super-resolution,'' in \emph{Proceedings of the IEEE conference on computer vision and pattern recognition}, 2018, pp. 2472--2481.

\bibitem{vdsr}
J.~Kim, J.~K. Lee, and K.~M. Lee, ``Accurate image super-resolution using very deep convolutional networks,'' in \emph{Proceedings of the IEEE conference on computer vision and pattern recognition}, 2016, pp. 1646--1654.

\bibitem{nlsa}
Y.~Mei, Y.~Fan, and Y.~Zhou, ``Image super-resolution with non-local sparse attention,'' in \emph{Proceedings of the IEEE/CVF Conference on Computer Vision and Pattern Recognition}, 2021, pp. 3517--3526.

\bibitem{he2022dster}
J.~He, Q.~Yuan, J.~Li, Y.~Xiao, X.~Liu, and Y.~Zou, ``Dster: A dense spectral transformer for remote sensing spectral super-resolution,'' \emph{International Journal of Applied Earth Observation and Geoinformation}, vol. 109, p. 102773, 2022.

\bibitem{ledig2017photo}
C.~Ledig, L.~Theis, F.~Husz{\'a}r, J.~Caballero, A.~Cunningham, A.~Acosta, A.~Aitken, A.~Tejani, J.~Totz, Z.~Wang \emph{et~al.}, ``Photo-realistic single image super-resolution using a generative adversarial network,'' in \emph{Proceedings of the IEEE conference on computer vision and pattern recognition}, 2017, pp. 4681--4690.

\bibitem{ddpm}
J.~Ho, A.~Jain, and P.~Abbeel, ``Denoising diffusion probabilistic models,'' \emph{Advances in neural information processing systems}, vol.~33, pp. 6840--6851, 2020.

\bibitem{srdiff}
H.~Li, Y.~Yang, M.~Chang, S.~Chen, H.~Feng, Z.~Xu, Q.~Li, and Y.~Chen, ``Srdiff: Single image super-resolution with diffusion probabilistic models,'' \emph{Neurocomputing}, vol. 479, pp. 47--59, 2022.

\bibitem{whl}
H.~Wu, N.~Ni, S.~Wang, and L.~Zhang, ``Conditional stochastic normalizing flows for blind super-resolution of remote sensing images,'' \emph{IEEE Transactions on Geoscience and Remote Sensing}, vol.~61, pp. 1--16, 2023.

\bibitem{miao2023dds2m}
Y.~Miao, L.~Zhang, L.~Zhang, and D.~Tao, ``Dds2m: Self-supervised denoising diffusion spatio-spectral model for hyperspectral image restoration,'' in \emph{Proceedings of the IEEE/CVF International Conference on Computer Vision}, 2023, pp. 12\,086--12\,096.

\bibitem{srcnn}
C.~Dong, C.~C. Loy, K.~He, and X.~Tang, ``Image super-resolution using deep convolutional networks,'' \emph{IEEE transactions on pattern analysis and machine intelligence}, vol.~38, no.~2, pp. 295--307, 2015.

\bibitem{enet}
M.~S. Sajjadi, B.~Scholkopf, and M.~Hirsch, ``Enhancenet: Single image super-resolution through automated texture synthesis,'' in \emph{Proceedings of the IEEE international conference on computer vision}, 2017, pp. 4491--4500.

\bibitem{rombach2022high}
R.~Rombach, A.~Blattmann, D.~Lorenz, P.~Esser, and B.~Ommer, ``High-resolution image synthesis with latent diffusion models,'' in \emph{Proceedings of the IEEE/CVF conference on computer vision and pattern recognition}, 2022, pp. 10\,684--10\,695.

\bibitem{xia2023diffir}
B.~Xia, Y.~Zhang, S.~Wang, Y.~Wang, X.~Wu, Y.~Tian, W.~Yang, and L.~Van~Gool, ``Diffir: Efficient diffusion model for image restoration,'' \emph{arXiv preprint arXiv:2303.09472}, 2023.

\bibitem{jiang2018deep}
K.~Jiang, Z.~Wang, P.~Yi, J.~Jiang, J.~Xiao, and Y.~Yao, ``Deep distillation recursive network for remote sensing imagery super-resolution,'' \emph{Remote Sensing}, vol.~10, no.~11, p. 1700, 2018.

\bibitem{lei2021transformer}
S.~Lei, Z.~Shi, and W.~Mo, ``Transformer-based multistage enhancement for remote sensing image super-resolution,'' \emph{IEEE Transactions on Geoscience and Remote Sensing}, vol.~60, pp. 1--11, 2021.

\bibitem{d2u}
Y.~Xiao, Q.~Yuan, K.~Jiang, J.~He, Y.~Wang, and L.~Zhang, ``From degrade to upgrade: Learning a self-supervised degradation guided adaptive network for blind remote sensing image super-resolution,'' \emph{Information Fusion}, vol.~96, pp. 297--311, 2023.

\bibitem{zs}
S.~Zhang, Q.~Yuan, J.~Li, J.~Sun, and X.~Zhang, ``Scene-adaptive remote sensing image super-resolution using a multiscale attention network,'' \emph{IEEE Transactions on Geoscience and Remote Sensing}, vol.~58, no.~7, pp. 4764--4779, 2020.

\bibitem{msdtgp}
Y.~Xiao, X.~Su, Q.~Yuan, D.~Liu, H.~Shen, and L.~Zhang, ``Satellite video super-resolution via multiscale deformable convolution alignment and temporal grouping projection,'' \emph{IEEE Transactions on Geoscience and Remote Sensing}, vol.~60, pp. 1--19, 2022.

\bibitem{hsenet}
S.~Lei and Z.~Shi, ``Hybrid-scale self-similarity exploitation for remote sensing image super-resolution,'' \emph{IEEE Transactions on Geoscience and Remote Sensing}, vol.~60, pp. 1--10, 2021.

\bibitem{chenshi}
S.~Chen, L.~Zhang, and L.~Zhang, ``Msdformer: Multiscale deformable transformer for hyperspectral image super-resolution,'' \emph{IEEE Transactions on Geoscience and Remote Sensing}, vol.~61, pp. 1--14, 2023.

\bibitem{mhan}
D.~Zhang, J.~Shao, X.~Li, and H.~T. Shen, ``Remote sensing image super-resolution via mixed high-order attention network,'' \emph{IEEE Transactions on Geoscience and Remote Sensing}, vol.~59, no.~6, pp. 5183--5196, 2021.

\bibitem{lgtd}
Y.~Xiao, Q.~Yuan, K.~Jiang, X.~Jin, J.~He, L.~Zhang, and C.-w. Lin, ``Local-global temporal difference learning for satellite video super-resolution,'' \emph{IEEE Transactions on Circuits and Systems for Video Technology}, pp. 1--16, 2023, doi: \href{https://doi.org/10.1109/TCSVT.2023.3312321}{10.1109/TCSVT.2023.3312321}.

\bibitem{yuanyuan1}
Q.~Li, M.~Gong, Y.~Yuan, and Q.~Wang, ``Symmetrical feature propagation network for hyperspectral image super-resolution,'' \emph{IEEE Transactions on Geoscience and Remote Sensing}, vol.~60, pp. 1--12, 2022.

\bibitem{yuanyuan3}
Q.~Li, Y.~Yuan, X.~Jia, and Q.~Wang, ``Dual-stage approach toward hyperspectral image super-resolution,'' \emph{IEEE Transactions on Image Processing}, vol.~31, pp. 7252--7263, 2022.

\bibitem{yuanyuan2}
Q.~Li, M.~Gong, Y.~Yuan, and Q.~Wang, ``Rgb-induced feature modulation network for hyperspectral image super-resolution,'' \emph{IEEE Transactions on Geoscience and Remote Sensing}, vol.~61, pp. 1--11, 2023.

\bibitem{lei2019coupled}
S.~Lei, Z.~Shi, and Z.~Zou, ``Coupled adversarial training for remote sensing image super-resolution,'' \emph{IEEE Transactions on Geoscience and Remote Sensing}, vol.~58, no.~5, pp. 3633--3643, 2019.

\bibitem{jiang2019edge}
K.~Jiang, Z.~Wang, P.~Yi, G.~Wang, T.~Lu, and J.~Jiang, ``Edge-enhanced gan for remote sensing image superresolution,'' \emph{IEEE Transactions on Geoscience and Remote Sensing}, vol.~57, no.~8, pp. 5799--5812, 2019.

\bibitem{usrgan}
J.~M. Haut, R.~Fernandez-Beltran, M.~E. Paoletti, J.~Plaza, A.~Plaza, and F.~Pla, ``A new deep generative network for unsupervised remote sensing single-image super-resolution,'' \emph{IEEE Transactions on Geoscience and Remote Sensing}, vol.~56, no.~11, pp. 6792--6810, 2018.

\bibitem{tu2022swcgan}
J.~Tu, G.~Mei, Z.~Ma, and F.~Piccialli, ``Swcgan: Generative adversarial network combining swin transformer and cnn for remote sensing image super-resolution,'' \emph{IEEE Journal of Selected Topics in Applied Earth Observations and Remote Sensing}, vol.~15, pp. 5662--5673, 2022.

\bibitem{liu2022diffusion}
J.~Liu, Z.~Yuan, Z.~Pan, Y.~Fu, L.~Liu, and B.~Lu, ``Diffusion model with detail complement for super-resolution of remote sensing,'' \emph{Remote Sensing}, vol.~14, no.~19, p. 4834, 2022.

\bibitem{han2023enhancing}
L.~Han, Y.~Zhao, H.~Lv, Y.~Zhang, H.~Liu, G.~Bi, and Q.~Han, ``Enhancing remote sensing image super-resolution with efficient hybrid conditional diffusion model,'' \emph{Remote Sensing}, vol.~15, no.~13, p. 3452, 2023.

\bibitem{anderson1982reverse}
B.~D. Anderson, ``Reverse-time diffusion equation models,'' \emph{Stochastic Processes and their Applications}, vol.~12, no.~3, pp. 313--326, 1982.

\bibitem{espcn}
W.~Shi, J.~Caballero, F.~Husz{\'a}r, J.~Totz, A.~P. Aitken, R.~Bishop, D.~Rueckert, and Z.~Wang, ``Real-time single image and video super-resolution using an efficient sub-pixel convolutional neural network,'' in \emph{Proceedings of the IEEE conference on computer vision and pattern recognition}, 2016, pp. 1874--1883.

\bibitem{kloeden1992stochastic}
P.~E. Kloeden, E.~Platen, P.~E. Kloeden, and E.~Platen, \emph{Stochastic differential equations}.\hskip 1em plus 0.5em minus 0.4em\relax Springer, 1992.

\bibitem{aid}
G.-S. Xia, J.~Hu, F.~Hu, B.~Shi, X.~Bai, Y.~Zhong, L.~Zhang, and X.~Lu, ``Aid: A benchmark data set for performance evaluation of aerial scene classification,'' \emph{IEEE Transactions on Geoscience and Remote Sensing}, vol.~55, no.~7, pp. 3965--3981, 2017.

\bibitem{dota}
G.-S. Xia, X.~Bai, J.~Ding, Z.~Zhu, S.~Belongie, J.~Luo, M.~Datcu, M.~Pelillo, and L.~Zhang, ``Dota: A large-scale dataset for object detection in aerial images,'' in \emph{Proceedings of the IEEE conference on computer vision and pattern recognition}, 2018, pp. 3974--3983.

\bibitem{dior}
K.~Li, G.~Wan, G.~Cheng, L.~Meng, and J.~Han, ``Object detection in optical remote sensing images: A survey and a new benchmark,'' \emph{ISPRS journal of photogrammetry and remote sensing}, vol. 159, pp. 296--307, 2020.

\bibitem{nwpu}
G.~Cheng, J.~Han, and X.~Lu, ``Remote sensing image scene classification: Benchmark and state of the art,'' \emph{Proceedings of the IEEE}, vol. 105, no.~10, pp. 1865--1883, 2017.

\bibitem{lpips}
R.~Zhang, P.~Isola, A.~A. Efros, E.~Shechtman, and O.~Wang, ``The unreasonable effectiveness of deep features as a perceptual metric,'' in \emph{Proceedings of the IEEE conference on computer vision and pattern recognition}, 2018, pp. 586--595.

\bibitem{dists}
K.~Ding, K.~Ma, S.~Wang, and E.~P. Simoncelli, ``Image quality assessment: Unifying structure and texture similarity,'' \emph{IEEE transactions on pattern analysis and machine intelligence}, vol.~44, no.~5, pp. 2567--2581, 2020.

\bibitem{ssim}
Z.~Wang, A.~C. Bovik, H.~R. Sheikh, and E.~P. Simoncelli, ``Image quality assessment: from error visibility to structural similarity,'' \emph{IEEE transactions on image processing}, vol.~13, no.~4, pp. 600--612, 2004.

\bibitem{salimans2016improved}
T.~Salimans, I.~Goodfellow, W.~Zaremba, V.~Cheung, A.~Radford, and X.~Chen, ``Improved techniques for training gans,'' \emph{Advances in neural information processing systems}, vol.~29, 2016.

\bibitem{niqe}
A.~Mittal, R.~Soundararajan, and A.~C. Bovik, ``Making a “completely blind” image quality analyzer,'' \emph{IEEE Signal processing letters}, vol.~20, no.~3, pp. 209--212, 2012.

\end{thebibliography}


%

\begin{IEEEbiography}[{\includegraphics[width=1in,height=1.25in,clip,keepaspectratio]{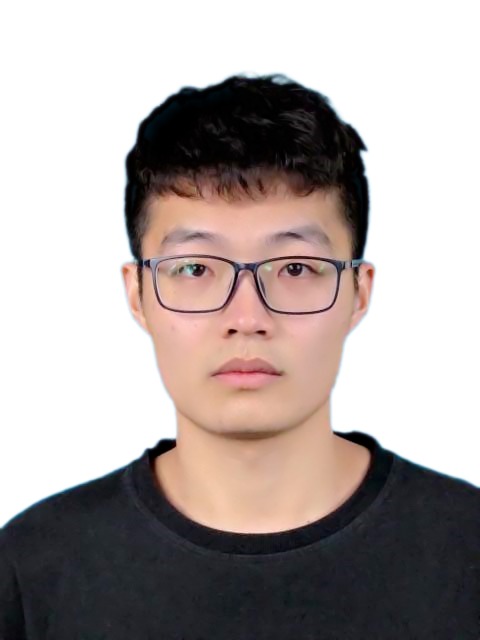}}]{Yi Xiao}
received the B.S. degree from the School of Mathematics and Physics, China University of Geosciences, Wuhan, China, in 2020. He is pursuing the Ph.D. degree with the School of Geodesy and Geomatics, Wuhan University, Wuhan.
\par His major research interests are remote sensing image super-resolution and computer vision. More details can be found at \url{https://xy-boy.github.io/}
\end{IEEEbiography}

\begin{IEEEbiography}[{\includegraphics[width=1in,height=1.25in,clip,keepaspectratio]{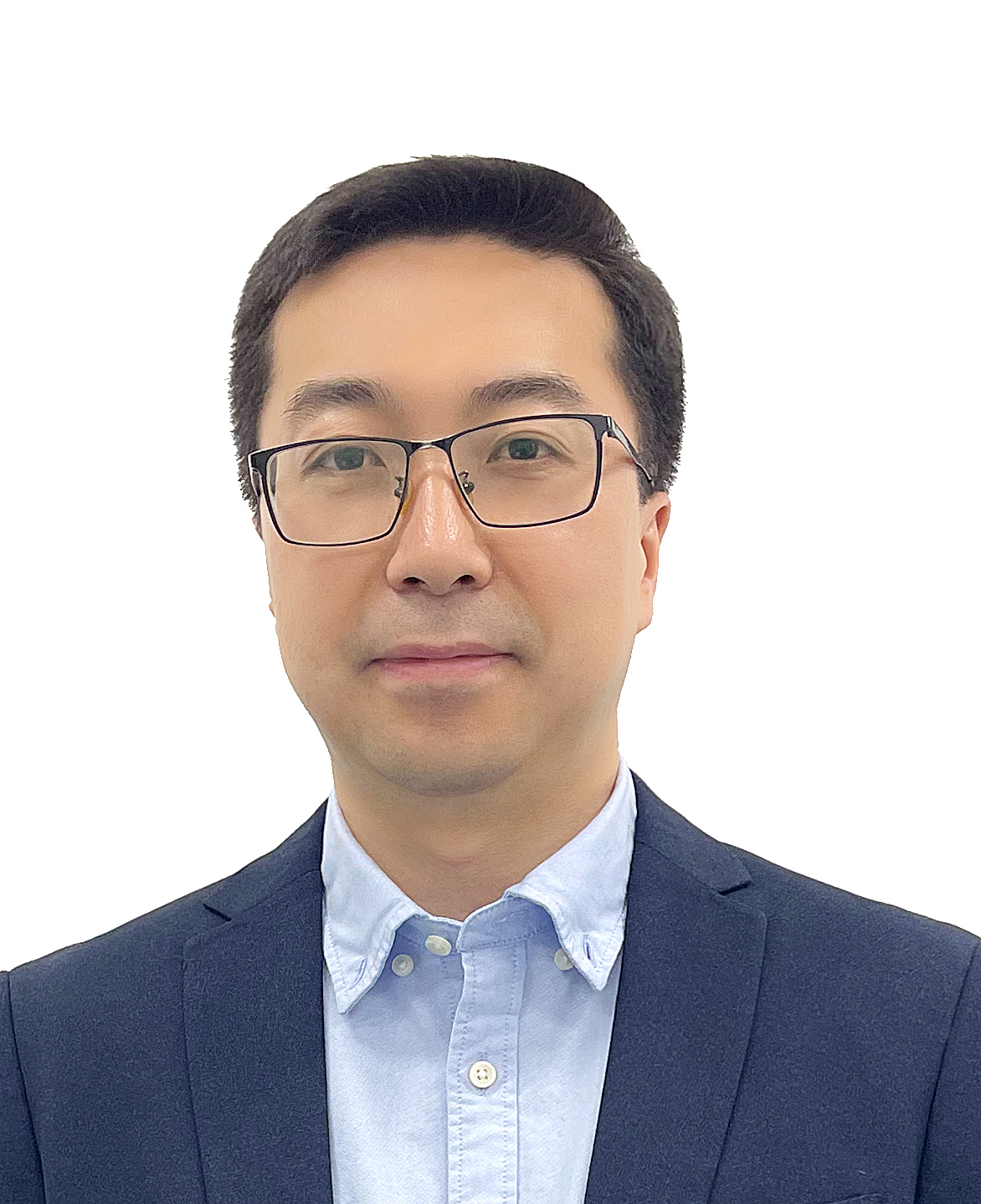}}]{Qiangqiang Yuan}
(Member, IEEE) received the B.S. degree in surveying and mapping engineering and the Ph.D. degree in photogrammetry and remote sensing from Wuhan University, Wuhan, China, in 2006 and 2012, respectively.
\par In 2012, he joined the School of Geodesy and Geomatics, Wuhan University, where he is a Professor. He has published more than 90 research papers, including more than 70 peer-reviewed articles in international journals, such as \emph{Remote Sensing of Environment, ISPRS Journal of Photogrammetry and Remote Sensing}, {\sc IEEE Transaction ON Image Processing}, and {\sc IEEE Transactions ON Geoscience AND Remote Sensing}. His research interests include image reconstruction, remote sensing image processing and application, and data fusion.
\par Dr. Yuan was a recipient of the Youth Talent Support Program of China in 2019, the Top-Ten Academic Star of Wuhan University in 2011, and the recognition of Best Reviewers of the IEEE GRSL in 2019. In 2014, he received the Hong Kong Scholar Award from the Society of Hong Kong Scholars and the China National Postdoctoral Council. He is an associate editor of 5 international journals and has frequently served as a referee for more than 40 international journals for remote sensing and image processing.
\end{IEEEbiography}

\begin{IEEEbiography}[{\includegraphics[width=1in,height=1.25in,clip,keepaspectratio]{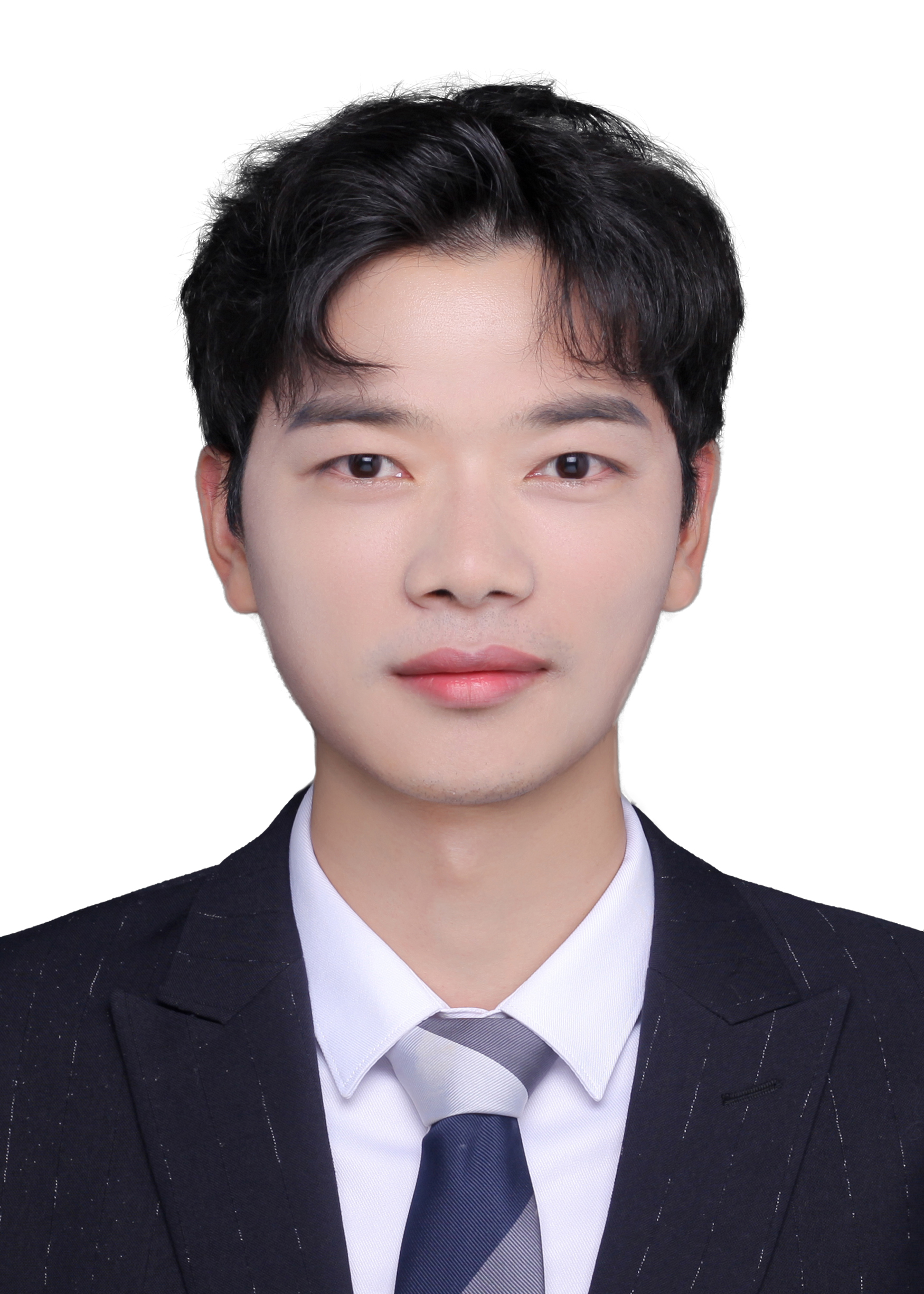}}]{Kui Jiang}
(Member, IEEE) received the Ph.D. degree in the school of Computer Science, Wuhan University, Wuhan, China, in 2022. He is currently an associate professor with the school of Computer Science and Technology, Harbin Institute of Technology, Harbin, China. He received the 2022 ACM Wuhan Doctoral Dissertation Award. His research interests include image/video processing and computer vision.
\end{IEEEbiography}

\begin{IEEEbiography}[{\includegraphics[width=1in,height=1.25in,clip,keepaspectratio]{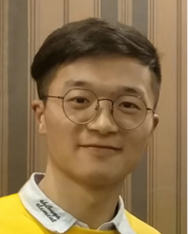}}]{Jiang He}
received the B.S. degree in remote sensing science and technology from faculty of geosciences and environmental engineering in Southwest Jiaotong University, Chengdu, China, in 2018. He is currently pursuing the Ph.D. degree in School of Geodesy and Geomatics, Wuhan University, Wuhan, China.
\par His research interests include hyperspectral super-resolution, image fusion, quality improvement, remote sensing image processing and deep learning.
\end{IEEEbiography}

\begin{IEEEbiography}[{\includegraphics[width=1in,height=1.25in,clip,keepaspectratio]{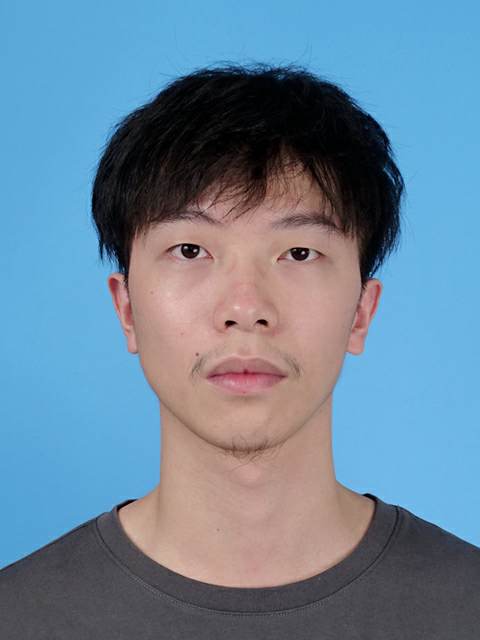}}]{Xianyu Jin}
received the B.S. degree in geodesy and geomatics engineering from Wuhan University, Wuhan, China, in 2019, where he is pursuing the M.S. degree with the School of Geodesy and Geomatics.
\par His research interests include video super-resolution, deep learning, and computer vision.
\end{IEEEbiography}

\begin{IEEEbiography}[{\includegraphics[width=1in,height=1.25in,clip,keepaspectratio]{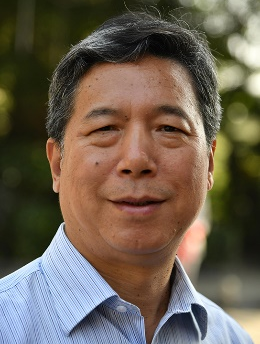}}]{Liangpei Zhang}
(Fellow, IEEE) received the B.S. degree in physics from Hunan Normal University, Changsha, China, in 1982, the M.S. degree in optics from the Xi’an Institute of Optics and Precision Mechanics, Chinese Academy of Sciences, Xi’an, China, in 1988, and the Ph.D. degree in photogrammetry and remote sensing from Wuhan University, Wuhan, China, in 1998.
\par He is currently a “Chang-Jiang Scholar” Chair Professor appointed by the Ministry of Education of China at the State Key Laboratory of Information Engineering in Surveying, Mapping, and Remote Sensing (LIESMARS), Wuhan University. He was a Principal Scientist for the China State Key Basic Research Project from 2011 to 2016 appointed by the Ministry of National Science and Technology of China to lead the Remote Sensing Program in China. He has published more than 700 research articles and five books. He is the Institute for Scientific Information (ISI) Highly Cited Author. He holds 30 patents. His research interests include hyperspectral remote sensing, high-resolution remote sensing, image processing, and artificial intelligence. 
\par Dr. Zhang is a fellow of the Institution of Engineering and Technology (IET). He was a recipient of the 2010 Best Paper Boeing Award, the 2013 Best Paper ERDAS Award from the American Society of Photogrammetry and Remote Sensing (ASPRS), and the 2016 Best Paper Theoretical Innovation Award from the International Society for Optics and Photonics (SPIE). His research teams won the top three prizes of the IEEE GRSS 2014 Data Fusion Contest. His students have been selected as the winners or finalists of the IEEE International Geoscience and Remote Sensing Symposium (IGARSS) Student Paper Contest in recent years. He is also the Founding Chair of the IEEE Geoscience and Remote Sensing Society (GRSS) Wuhan Chapter. He also serves as an associate editor or an editor for more than ten international journals. He is also serving as an Associate Editor for the {\sc IEEE Transactions ON Geoscience AND Remote Sensing}.
\end{IEEEbiography}

\end{document}